\documentclass[aps,prx,twocolumn,floatfix,superscriptaddress]{revtex4-2}

\usepackage{amsfonts,mathrsfs,amsmath,amsthm,amssymb}
\usepackage{bm,times}
\usepackage{graphicx,epsfig}
\usepackage[colorlinks=true,breaklinks=true,linkcolor=blue,citecolor=blue,urlcolor=blue]{hyperref}
\usepackage{booktabs}
\def \non {\nonumber}

\newcommand{\be}{\begin{eqnarray}}
\newcommand{\ee}{\end{eqnarray}}

\newcommand{\zenodoconcept}{10.5281/zenodo.23794180}
\newcommand{\smconcept}{10.5281/zenodo.23914166}

\makeatletter
    
    \newcommand{\Rmnum}[1]{\expandafter\@slowromancap\romannumeral #1@}
\makeatother

\begin{document}
\title{Probing Coherent Noise with Stabilizer Codes:\\ Identifiability, Ambiguity, and Decoder Priors}
\author{Long Huang}
\email{huangl@sicnu.edu.cn}
\affiliation{College of Physics and Electronic Engineering, Sichuan Normal University, Chengdu 610101, China}

\begin{abstract}
Coherent noise is erased by Pauli twirling and beyond process tomography. We turn measurement-free quantum error correction---syndrome-free coherent feedback---into a probe: bit-flip and phase-flip codes polarize a few-qubit coherent error into single-logical-qubit channels, and seven identify an eleven-parameter three-qubit generator. Three references are read with the spectators in $|0\ldots0\rangle$, two with the spectators in $|+\ldots+\rangle$, and two are the code channels; each one single-qubit tomography, so the fingerprint is $112$ components at $84$ settings. An exact rational-arithmetic rank certificate proves local injectivity on a dense open subset, with the smallest singular value between $1.75$ and $1.83$ over $500$ sampled points, so conditioning is uniform, not merely finite. The $|0\ldots0\rangle$ references alone are not globally injective: a compensating rotation combined with the site exchange gives exact collisions; we classify them completely: the reflection is the only nontrivial permutation symmetry of the $n$-qubit generator. The $|+\ldots+\rangle$ references remove that ambiguity by construction and separate every nontrivial mirror pair. Because they dephase in proportion to the spectator magnetization, they also make visible the out-of-family Pauli directions---including the common $Z_{1}Z_{3}$ crosstalk---that a $|0\ldots0\rangle$ reference absorbs at first order. A Cram\'er--Rao analysis reaches $0.01$~rad per parameter with $1.5\times10^{3}$ shots per component, $1.3\times10^{5}$ events at twelve settings per channel, and a per-gate calibration allowance of $1.6\times10^{-2}$, so the probe is shot-noise limited. In a distance-three surface code the generator gives a $14.8$-fold encoding gain per site and a further $1.24$-fold ($1.21$-fold from an estimate) from the recovered correlations.
\end{abstract}

\date{\today}

\maketitle
\emph{Introduction.}---Coherent noise differs from stochastic Pauli noise in three essential ways~\cite{Li,Harrington,Lupke}: first, coherent errors preserve relative phases among different error paths, allowing amplitudes to interfere; second, they accumulate over QEC rounds, producing residual logical coherence; third, for small unitary errors their worst-case diamond distance scales with the square root of the average gate infidelity under stated conventions, making Pauli-error-rate estimates overly optimistic. Typical physical sources include control-pulse under-rotations, residual two-qubit $ZZ$ coupling, slowly drifting control fields, and non-Markovian bath correlations. Recovering the parameters of such a physical generator serves a purpose beyond estimating the error rate of an existing circuit: it supports attribution of couplings to hardware locations, propagation of noise under new gate sequences, and construction of circuit-dependent priors for decoders.

Full quantum process tomography, however, requires reconstructing $16^{n}$ real parameters and becomes intractable beyond a few qubits. Recent work has developed several scalable forms of compression. Averaged circuit eigenvalue sampling (ACES) estimates gate-resolved Pauli probabilities and averaged spatial correlations in large syndrome-extraction circuits~\cite{ACES}, and can calibrate correlated matching decoders with substantial improvements in error suppression~\cite{NoiseAware}. Decoding-graph re-weighting (DGR) instead adapts edge and pair weights directly from decoding statistics~\cite{DGR}, while syndrome-only estimation reconstructs detector graphs and hypergraphs, including multi-body fault correlations~\cite{Takou,Cheng}. Complementary process-level approaches include compressed-sensing tomography for sparse or low-rank $\chi$ matrices~\cite{Shabani}, low-degree channel tomography for coherent and correlated gate-layer noise with logarithmic overhead~\cite{Crupi}, and spacetime classical shadows that distinguish crosstalk from genuine non-Markovianity~\cite{White}.

These advances also clarify the division of labor. For immediate calibration of a decoder on an existing circuit, Pauli statistics and detector-level correlations are often sufficient, and full physical parameter recovery is unnecessary. ACES and DGR are particularly effective in this reactive or circuit-specific setting. A compressed Pauli model, however, does not by itself identify the Hamiltonian term responsible for an observed correlation, nor does it uniquely determine how that correlation will change under a different control sequence. Physical parameter recovery addresses this complementary diagnostic and predictive task; it should augment, rather than replace, Pauli-statistics-based calibration.

The cheapest alternative for a small model is direct targeted spectroscopy, and we compare against it explicitly rather than setting it aside. Rabi, Ramsey and phase tomography give the three single-qubit triples $(\alpha_i,\beta_i,\delta_i)$, and echo or $ZZ$-interference sequences give the two couplings, so for eleven parameters this is cheaper per circuit than any code-based probe; two-qubit process tomography, by contrast, carries $256$ real components. Three differences motivate the present design. First, what is measured differs: idle spectroscopy characterizes a qubit that is not being operated, whereas the channel entering a decoder is the one acting inside the encoding, syndrome, and recovery sequence, which includes coherent error from those gates. Second, the fingerprint is compressive: $48$ single-qubit components reach the same eleven-parameter local identifiability that two-qubit tomography reaches with $256$, and they require single-qubit measurements only, so no mid-circuit readout or two-qubit state tomography is needed. Third, targeted spectroscopy returns point estimates without a channel-level Fisher information, whereas the fingerprint supplies the Jacobian, conditioning, and shot-count analysis on which the resource comparison below rests. The three approaches are therefore complementary: spectroscopy for a first physical picture, two-qubit tomography for a reference channel, and polarization fingerprints when the object of interest is the effective channel inside an error-correction cycle together with its uncertainty.

Here we pursue a complementary route: using the encoding itself as the probe. Two elementary stabilizer codes---the bit-flip and phase-flip codes---polarize a coherent error toward the Pauli-$Z$ and Pauli-$X$ axes under measurement-free QEC, and different concatenation orders give distinguishable effective channels~\cite{Huang}. We call the stacked Pauli transfer matrix of these single-logical-qubit channels a \emph{fingerprint}: $16$ real entries per channel, each obtained by single-qubit process tomography. Two code channels are the probe; five product-state reference channels, read with the spectator qubits in $|0\ldots0\rangle$ or in $|+\ldots+\rangle$, are the calibration around it, and the seven together make a $112$-component fingerprint at $84$ settings.

The central question we answer is what such a probe can and cannot learn. On the one hand, an exact rational-arithmetic rank certificate, evaluated at a rational point inside $\Theta$, proves that the $48\times11$ Jacobian of the code core has full column rank on a dense open subset of the compact domain $\Theta$, so the map is locally injective outside a proper analytic variety; $500$ random Jacobians confirm full rank throughout, with $s_{\min}$ between $1.75$ and $1.83$---uniform conditioning, not merely finite. On the other hand the map is \emph{not} globally injective as soon as the references are all of one kind. The code channels are exactly invariant under the site reflection $1\leftrightarrow3$, and a $|0\ldots0\rangle$ reference, although it improves conditioning and separates generic mirror pairs, fails on the codimension-three locus $U_3e^{-ibZ}=U_1e^{-iaZ}$: we exhibit two regular in-domain points with identical $|0\ldots0\rangle$-frame fingerprints but physically inequivalent unitaries. We classify this reflection-generated ambiguity completely. The reflection is the only nontrivial permutation symmetry of the $n$-qubit generator (Proposition~1), so the classification is a structural statement about a class of models rather than a property of one example. The ambiguity is then removed by construction rather than by an appended measurement: with the spectators in $|+\rangle$ the reference becomes a dephasing channel whose rate is the spectator magnetization, which resolves the product $U_qe^{-ia_qZ}$ into its factors and separates every nontrivial mirror pair (Lemma~1, Theorem~4, Corollary~4). A two-qubit marginal $\mathcal R_{12}$ closes the same sector at the price of a two-qubit tomography per calibration session (Theorem~3); we price it but do not need it, and the twelve-setting route costs three hundred times fewer events once each route is given the shots it needs. Lemma~1 is stated after Theorem~3 because it is used only by Theorem~4, and because the product form it describes is what makes the $|+\rangle$ reference the natural comparison with the marginal.

The choice of seven channels is a design decision with a measurable answer rather than a convention. A joint enumeration of all $262\,080$ assemblies available to this model selects the set used here as the cheapest that retains the code diagnostic, and the five references alone, without the code channels, already attain rank $11$; the code channels raise $s_{\min}$ from $1.06$ to $1.76$ and are carried because they are the object of the measurement. Relative to the companion work~\cite{Huang}, which proves the polarization redistribution theorem and the distinguishability of concatenation orders, the present paper contributes the identifiability theory (Theorems~1--4, Lemma~1, Proposition~1 and Corollary~4, with the complete reflection classification), the error and shot budgets, and the decoder-prior interface; Observation~1 invokes Ref.~\cite{Huang} only for the polarization tendency and adds no claim to it.

A Cram\'er--Rao analysis, checked at the level of individual measurement settings, reaches $0.01$~rad per parameter at a representative point with $1.5\times10^{3}$ shots per component, that is $1.3\times10^{5}$ events at twelve settings per channel and seven channels, or $6.8\times10^{4}$ events in the measurement-level calculation; a literal reading of the per-component convention gives $1.7\times10^{5}$, so the two conventions differ by a factor $2.5$ and not by an order of magnitude. Because the conditioning is uniform the sample does not degrade this: the worst of the $500$ points needs $1.6\times10^{3}$ shots. Propagating calibration uncertainty through the recursion bounds the per-gate residual at $1.6\times10^{-2}$ (stochastic) or $7.7\times10^{-3}$ (coherent) for the fingerprint RMS to stay below the shot noise, so a per-gate residual of $10^{-3}$ leaves the probe shot-noise limited; a conservative worst-direction sufficient condition for $0.01$~rad is $2.6\times10^{-4}$. Finally, propagating the generator into a nine-qubit rotated surface code with ideal syndrome extraction reduces the logical infidelity from $2.66\times10^{-3}$ to $1.79\times10^{-4}$ when the generator is known exactly, and retaining the correlations gives a further $1.239$-fold gain (the $1.24$ quoted above, rounded) over a decoder built from identical single-qubit marginals---$1.21$ when the prior is rebuilt from an estimate at the stated precision, so the quoted $1.21$ is the correlation gain itself and not a product with $1.239$.

What makes the output reusable is that each fingerprint is a single-logical-qubit channel, so the settings per fingerprint are fixed while the number of fingerprints is set by $4n-1$; the model class, not a sparsity penalty, is the structural assumption.

\begin{figure*}[tb]
\centering
\includegraphics[width=0.72\textwidth]{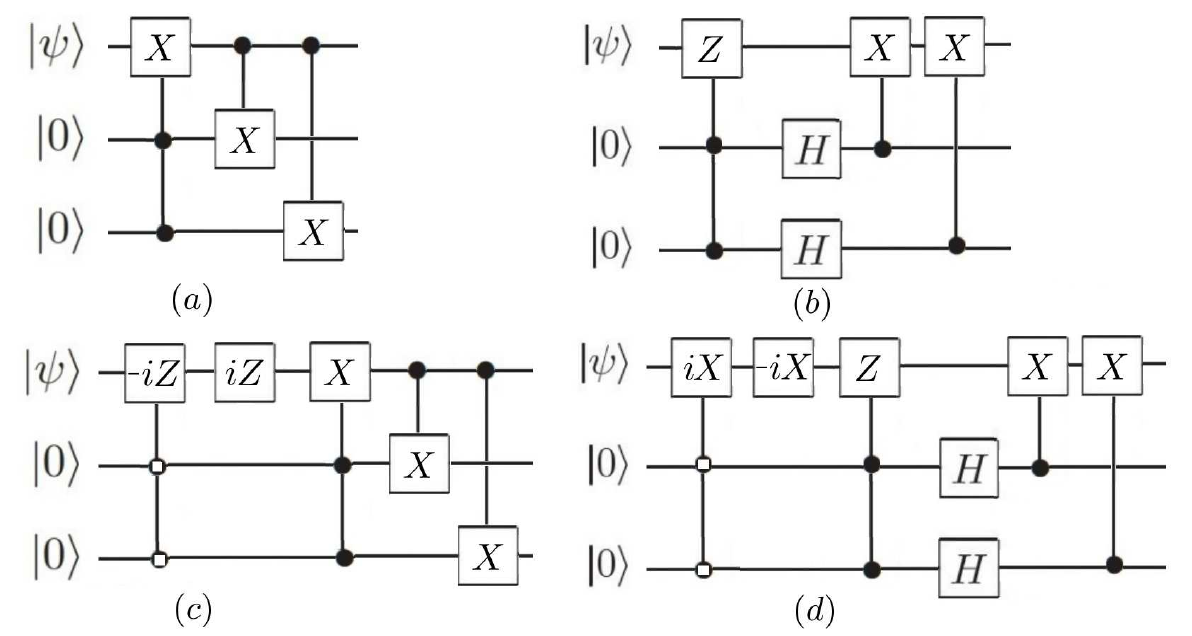}
\caption{(a) Encoding circuit for the bit-flip code.
 (b) Encoding circuit for the phase-flip code. (c) Measurement-free QEC circuit using the code in Eq.~(\ref{e1}).
 (d) Measurement-free QEC circuit using the code in Eq.~(\ref{e2}).}\label{figure1}
\end{figure*}

\emph{Three-qubit code examples.}---We use two elementary three-qubit codes as polarization probes rather than as full fault-tolerant codes~\cite{Shor,Steane,Laflamme,Bennett}. After QEC, the bit-flip code in Eq.~(\ref{e1}) polarizes the effective channel toward Pauli-$Z$, whereas the phase-flip code in Eq.~(\ref{e2}) polarizes it toward Pauli-$X$. The bit-flip code has stabilizer generators $\langle Z_1Z_2, Z_2Z_3\rangle$ and logical basis states
\be
\label{e1}
|0_\mathcal{L}\rangle=|000\rangle,\non\\
|1_\mathcal{L}\rangle=|111\rangle.
\ee
Using this code, the state $|\psi\rangle=a|0\rangle+b|1\rangle$ is encoded as $|\tilde{\psi}\rangle=a|000\rangle+b|111\rangle$. The corresponding measurement-free QEC protocol is shown in Figs.~\ref{figure1}(a) and \ref{figure1}(c).
The phase-flip code has stabilizer generators $\langle X_1X_2, X_2X_3\rangle$ and logical basis states
\be
\label{e2}
|0_\mathcal{L}\rangle=\frac{1}{2}[|000\rangle+|011\rangle+|101\rangle+|110\rangle],\non\\
|1_\mathcal{L}\rangle=\frac{1}{2}[|111\rangle+|100\rangle+|010\rangle+|001\rangle].
\ee
In this case, $|\psi\rangle$ is encoded as $|\tilde{\psi}\rangle=a[|000\rangle+|011\rangle+|101\rangle+|110\rangle]+b[|111\rangle+|100\rangle+|010\rangle+|001\rangle]$. The corresponding measurement-free QEC protocol is shown in Figs.~\ref{figure1}(b) and \ref{figure1}(d).
Both protocols use measurement-free QEC; the code structure induces polarization rather than providing full fault tolerance.

\emph{Polarization Effects and Identifiability of the Sample Codes.}---For an arbitrary quantum channel $\mathcal{E}$, we define the effective channel $\mathcal{E}_{\mathrm{eff}}$ as the logical channel obtained after encoding, propagation through $\mathcal{E}$, and measurement-free correction. Numerical simulation shows that the bit-flip code polarizes the effective channel toward the Pauli-$Z$ direction, while the phase-flip code polarizes it toward the Pauli-$X$ direction, in agreement with the analytic redistribution theorem of Ref.~\cite{Huang}. We formalize this as follows.

\textbf{Observation 1 (Polarization tendency for the specified unitary family)}. Consider the small-angle unitary family specified below, or a perturbative expansion of a channel around it. After one round of measurement-free QEC with the bit-flip code, the leading nontrivial Pauli-transfer-matrix block of the effective channel is dominated by the $Z$ axis; after the phase-flip code, it is dominated by the $X$ axis. If the noise contains coherent components along both $X$ and $Z$, the two codes emphasize orthogonal axes at the effective-channel level. This is a perturbative tendency for the stated family, not a universal theorem for an arbitrary channel containing coherent components; a rotation about an already-corrected axis requires a separate order-by-order treatment. Appendix E makes this quantitative: for a rotation about $Z$ the bit-flip channel responds as $18\varepsilon^{2}$, while the phase-flip channel responds only as $6\varepsilon^{4}$, and the two axes are interchanged for a rotation about $X$.

\textbf{Observation 2 (Fingerprint distinguishability)}. When the two codes are concatenated, the two orders generally yield different effective channels:
$$\mathcal{E}_{\mathrm{eff}}^{(ZX)}\neq\mathcal{E}_{\mathrm{eff}}^{(XZ)}\quad\text{(in general)},$$
where the superscript denotes the order, e.g., bit-flip first and phase-flip second $(XZ)$. This is the natural first guess at additional measured components, and we record it because Ref.~\cite{Huang} proves the distinguishability. It does not by itself create a new parameter fiber: once the first-level logical channels are fixed, every later logical channel is a deterministic function of them, so the additional components are statistically correlated repeats rather than new views. The measurements of Sec.~IV put a number on that: the four concatenated blocks cost $85$ gate uses per repetition and $1.7\times10^{8}$ events for a precision the selected references reach with $17$ gate uses and $1.3\times10^{5}$ events.

Let the coherent noise be represented by a parameterized family $\{\mathcal{E}_\theta:\theta\in\Theta\}$, where $\Theta$ is compact. If there exists an integer $L$ such that the map
$$\Phi_L:\;\Theta\to \left(\mathcal{E}_{\mathrm{eff}}^{(p_1)},\,\mathcal{E}_{\mathrm{eff}}^{(p_2)},\,\dots,\,\mathcal{E}_{\mathrm{eff}}^{(p_{M_L})}\right)$$
is locally injective outside a measure-zero exceptional subset of $\Theta$, then the effective-channel data determine the parameters in a neighborhood. A global statement requires an additional argument and is not assumed here. Concatenation depth does not supply new projections needed for identifiability. If $F(\theta)$ denotes a set of channels and $G$ the fixed encoding-correction map defining a deeper level, then the observation is $M(\theta)=(F(\theta),G(F(\theta)))$, so $M(\theta)=M(\theta')$ if and only if $F(\theta)=F(\theta')$ and $DM=[I;DG]DF$: a deeper level can add statistically independent measurements but cannot remove a fiber or add a locally identifiable direction. Its possible precision benefit must therefore be assessed at equal gate count, noise-primitive use, and calibration cost, and for the measurements of Appendix~D it is negative on all three. The identifying content has to come from channels read directly rather than through a code, and it has to come in more than one product frame, because a $|0\ldots0\rangle$ frame measures only the products $U_qe^{-ia_qZ}$, whose two-parameter family of kernel directions is the subject of Lemma~1. The selected fingerprint therefore keeps the two code channels---they are the object of the measurement---and surrounds them with five product references in two frames. If the actual noise falls outside the pre-specified family, the protocol returns the closest in-family approximation, and model-selection criteria must be applied---but not to the fit residual. At $\theta_{\rm res}$ no direction of the fifty-two out-of-family Pauli terms is absorbed below $8\%$ of its response (the weakest is $\rho=0.11$); over the fifteen sampled points of Appendix~F the smallest is $0.053$, so a good $\chi^{2}$ is still not evidence that the family is right: a term absorbed at $\rho=0.05$ leaves a residual of $1.5\times10^{-4}$ in fingerprint norm, two orders of magnitude below the per-component shot noise at the operating budget, which is why the check has to be an added nuisance column rather than a residual test, and why it belongs in the protocol rather than in the post-hoc analysis. What remains is a bounded sensitivity floor rather than a blind spot: the stochastic budget of Appendices~A and~G bounds the model error of any term that evades the fit, and Appendix~F quantifies what a term of given size does to the decoder gain.

\textbf{Protocol steps}. The complete protocol is as follows.

\emph{(i)} Perform single-qubit quantum process tomography and convert the result to the real PTM $\mathcal R'_0$ using the convention in Appendix C. Based on $\mathcal R'_0$, pre-specify a candidate coherent-noise family $\mathcal{E}_0$ applicable to three qubits.

\emph{(ii)} Apply one level of measurement-free QEC. Numerically, scan parameters within $\mathcal{E}_0$ to generate effective-channel datasets $\mathcal R_{1X}$ and $\mathcal R_{1Z}$; experimentally, perform single-qubit QPT on the effective channels and convert the results to $\mathcal R'_{1X}$ and $\mathcal R'_{1Z}$. Match the experimental data against the datasets using a minimum-distance criterion, screen candidate models, and refine the noise model to obtain $\mathcal{E}_1$.

\emph{(iii)} Repeat the single-qubit QPT of step (i) on sites $2$ and $3$ (spectators in $|00\rangle$) and on sites $1$ and $3$ with the two spectator qubits prepared in $|+\rangle$. Together with the two channels of step (ii) this gives the seven-channel fingerprint, and the eleven parameters are recovered by Gauss--Newton inversion of $\mathbf M(\theta)$.

The recovered object is a parameterized physical generator, not only a static Pauli error table. For a subsequent gate schedule $\mathcal C$, the same parameters can be propagated through $\mathcal C$ to predict the transformed noise channel, including changes in the Pauli type and in interference between error paths. Decoder weights for that schedule can then be obtained by Pauli projection, by an amplitude-aware likelihood model, or by a hybrid of the two. This predictive step is what distinguishes a reusable physical model from a decoder graph fitted only to past syndrome or matching statistics. A pulse-level implementation is given in the Supplemental Material~\cite{SM}: the circuit library, the twelve tomography settings per channel, readout correction, null-channel deconvolution of state-preparation and measurement error, the interleaved acquisition schedule, and the expected precision at a fixed shot budget. We demonstrate this propagation numerically for a distance-three rotated surface code below and in Appendix B; circuit-level propagation with noisy syndrome extraction remains a separate experimental task.

\textbf{Setup.} Let the physical coherent noise be a unitary channel
\be
\mathcal{E}_\theta(\rho)=K_\theta\rho K_\theta^\dagger,
\ee
\be
K_\theta=\Bigl(\bigotimes_{i=1}^{3}e^{-i(\alpha_i X^{(i)}+\beta_i Y^{(i)}+\delta_i Z^{(i)})}\Bigr)
e^{-i(\gamma_{12}Z_1Z_2+\gamma_{23}Z_2Z_3)}.\non\\
\ee
with parameter domain $\Theta=[-0.1,0.1]^9\times[-0.05,0.05]^2$, stacked in the block order $(\alpha_1,\alpha_2,\alpha_3,\beta_1,\beta_2,\beta_3,\delta_1,\delta_2,\delta_3,\gamma_{12},\gamma_{23})$ used throughout and in Appendix~C. The observation map is assembled from seven single-qubit channels, each obtained by one single-qubit process tomography: three product references $\Lambda_q^{\rm Z}(\theta)$, the reduced channel on site $q$ when $K_\theta$ acts on $\rho_q\otimes|00\rangle\langle00|$, with $\Lambda_1^{\rm Z}$ the physical channel of step (i); the two measurement-free code channels $\mathcal R_X,\mathcal R_Z$ of the bit-flip and phase-flip codes, from step (ii); and two further references $\Lambda_q^{\rm X}$, $q=1,3$, in which the two spectator qubits are prepared in $|+\rangle$ (Lemma~1 below). Thus
$$\mathbf M(\theta)=\bigl(\Lambda_1^{\rm Z},\Lambda_2^{\rm Z},\Lambda_3^{\rm Z},\mathcal R_X,\mathcal R_Z,\Lambda_1^{\rm X},\Lambda_3^{\rm X}\bigr),$$
where each entry is the real $4\times4$ Pauli transfer matrix (PTM) of the corresponding single-qubit channel, flattened row by row in the Pauli order $(I,X,Y,Z)$, so $\mathbf M:\Theta\to\mathbb R^{112}$: $16$ components per channel at twelve settings per channel, $84$ settings in all. All singular values and calibration norms below refer to the PTM convention of Appendix~C. This particular set of seven is not a guess. The companion paper~\cite{Companion} enumerates all $4095$ subsets of the twelve channel types available to this model against all $64$ recovery tables, $262\,080$ assemblies in total, and finds that among the assemblies that retain the code sub-block---which is what makes a wrong recovery table visible, see Appendix~F---this one minimizes the event count at fixed per-channel precision: $1.3\times10^{5}$ events and $17$ gate uses per repetition, against $1.7\times10^{8}$ events and $85$ gate uses for the two-level concatenation used in earlier versions of this protocol. The five references alone, without the code channels, have the same rank and a slightly worse $s_{\min}$ ($1.06$ against $1.76$), so the code channels are carried because they are the object of the measurement, not because identifiability needs them. The second-level block of the earlier design has an exact mirror degeneracy and contributes almost nothing at first order, which is why replacing it by two $|+\rangle$ references is a strict improvement on both the conditioning and the cost.

The distinction between the PTM and the process matrix matters throughout: every singular value, and every calibration norm, below refers to the convention of Appendix~C. The code channels alone have an exact mirror degeneracy. The product references improve the conditioning and separate generic mirror pairs, but a $|0\ldots0\rangle$ reference does not remove every mirror ambiguity; an exact counterexample is given after Theorem 1, and the $|+\rangle$ references are what close it (Theorem 4). Theorem 1 establishes generic local identifiability only.

\textbf{Theorem 1 (Generic local identifiability).} Let $\mathbf M$ be the $112$-component analytic fingerprint above. There exists a proper real-analytic subvariety $V\subset\Theta$, of Lebesgue measure zero, such that $\operatorname{rank}J(\theta)=11$ for every $\theta\in\Theta\setminus V$. Consequently $\mathbf M$ is locally injective on $\Theta\setminus V$. No statement about the global cardinality or finiteness of a fiber is implied.

\textbf{Proof}. Every PTM entry is analytic in the angular parameters, so $\mathbf M$ is real analytic. The $48$ rows of the core submap $F=(\Lambda_1^{\rm Z},\mathcal R_X,\mathcal R_Z)$ are $48$ of the $112$ rows of $\mathbf M$, and adjoining rows cannot lower a rank, so $\operatorname{rank}J\ge\operatorname{rank}DF$ everywhere. It is therefore enough to certify rank $11$ for the $48\times11$ core Jacobian at a single point in the interior of $\Theta$, which Appendix~C does in exact rational arithmetic: the determinant of an $11\times11$ minor is a nonzero rational number. The rank-deficient locus $V=\{\theta:\operatorname{rank}J(\theta)<11\}$ is contained in the zero set of $\det(J^TJ)$, a proper analytic subvariety once one nonzero minor is exhibited; it has measure zero, and off it the inverse function theorem gives a local inverse, hence local injectivity. Numerically the origin is not in $V$ either---the full Jacobian is regular there, with $s_{\min}=1.748$, the low end of the sampled range---although the $48$-row core degenerates at the same point to rank $5$. This is an observation and not part of the certificate, which needs only one regular point, and Appendix~C supplies it. The mechanism is worth recording, because it is why the core cannot be the whole probe: the kernel of a single $|0\ldots0\rangle$ reference mixes the two couplings with the $Z$ angles of that site, and the $|+\rangle$ references see the angle part at first order even where their $\gamma$-derivative vanishes as $\sin^{2}\gamma$. The compactness of $\Theta$ and analyticity of $\mathbf M$ do not by themselves imply that a fiber is finite, so no such conclusion is drawn. $\square$

The certification is by exact evaluation rather than by numerical sampling. Write each single-qubit rotation in the quaternion chart $(a_i,b_i,c_i,d_i)$ with $a_i=\sqrt{1-b_i^2-c_i^2-d_i^2}$ and each $ZZ$ coupling as $t=\tan\gamma$. In this chart the fingerprint entries are algebraic functions whose values and first derivatives can be propagated by exact forward-mode differentiation, and the rank of the differential is coordinate-independent. At the rational point given in Appendix~C, which lies in the interior of $\Theta$---so the nonzero minor is exhibited inside the domain that the numerical work and the physical protocol actually use, and no analytic continuation is invoked---the core Jacobian has exact rank $11$; the determinant of the $11\times11$ minor recorded there is a nonzero rational number, and its small magnitude is a small-angle effect that plays no role in the argument, which uses exactness rather than size. Appendix~C also records the coordinate conventions, row ordering, the executable checks, and the certificate in the form originally computed for a $64$-row completion of the core.

\textbf{Numerical symmetry test.} For each of the $2^{11}-1$ nontrivial sign flips $s\in\{\pm1\}^{11}$ we evaluate $\Delta_s(\theta)=\mathbf M(\theta)-\mathbf M(s\theta)$ at the certified point of Appendix~C, using an implementation whose exact mirror covariance in the $|0\ldots0\rangle$ frame is reproduced to $2.2\times10^{-16}$. Every flip gives a nonzero response; the smallest is $\max|\Delta_s|=1.30\times10^{-1}$, the $\gamma_{12}$ flip, the median over all $2047$ flips is $0.41$, and every single flip exceeds $1.29\times10^{-1}$. Sign-flip collisions are therefore absent at the tested point. The test cannot detect a compensating rotation combined with a site permutation, which is why the classification below is needed.

\textbf{Remark (Mirror degeneracy and the exact compensated collision).} The noise family, both codes, and the standard measurement-free recovery are covariant under the reflection exchanging qubits $1$ and $3$. Let $S$ act by $(\alpha_1,\beta_1,\delta_1)\leftrightarrow(\alpha_3,\beta_3,\delta_3)$ and $\gamma_{12}\leftrightarrow\gamma_{23}$, with qubit $2$ fixed. Then $U_{\mathrm{swap}}K_\theta U_{\mathrm{swap}}^\dagger=K_{S\theta}$, and every code-based channel satisfies $\mathcal R_c(S\theta)=\mathcal R_c(\theta)$ at every concatenation depth. The code-only fingerprint is therefore exactly twofold degenerate, and no finite concatenation depth removes this symmetry.

The physical reference channel is

\[
\Lambda_1^{\rm Z}(\theta)=\operatorname{Ad}_{U_1e^{-iaZ}},
\qquad
\Lambda_1^{\rm Z}(S\theta)=\operatorname{Ad}_{U_3e^{-ibZ}},
\]

because the other two qubits are initialized in $|00\rangle$. Consequently $\Lambda_1^{\rm Z}$ separates a generic mirror pair when the outer rotations differ, but it does not separate all of them. If $a\neq b$ and

\[
U_3=U_1e^{-i(a-b)Z},
\]

then $U_3e^{-ibZ}=U_1e^{-iaZ}$ and the physical channels coincide. Reflection invariance of the codes then gives an exact collision of the $|0\ldots0\rangle$-frame submap, that is of the five blocks built from $\Lambda_q^{\rm Z}$ and the two codes,

\[
\boxed{\mathbf M_Z(\theta)=\mathbf M_Z(S\theta),\qquad \theta\neq S\theta,}
\]

while the two $|+\rangle$ references do not collide (Theorem~4). Appendix C gives two rational parameter points satisfying this relation strictly inside $\Theta$, with exact equality of all $80$ components of $\mathbf M_Z$. Both points have Jacobian rank $11$; the corresponding three-qubit unitaries are not related by a global phase. Hence a unitary reference held on $|0\rangle$ improves conditioning and separates generic mirror pairs, but it cannot restore global resolution on its own, and a sign-flip test cannot detect this compensating rotation combined with a site permutation. This is the precise sense in which the reflection ambiguity is a degeneracy of the product frame, and the reason the selected fingerprint carries references in two frames.

\textbf{Theorem 2 (Classification of reflection-generated ambiguities).} For every $\theta\in\Theta$, all code-based channels are invariant under $S$. The $|0\ldots0\rangle$-frame submap satisfies $\mathbf M_Z(\theta)=\mathbf M_Z(S\theta)$ if and only if the physical channels agree, which for two points in the small-angle domain is equivalent to
\[
U_3e^{-ibZ}=U_1e^{-iaZ}.
\]
The collision locus $C_{\rm mir}$ defined by this equation is a codimension-three real-analytic submanifold of dimension $8$: it is parametrized by $(U_1,U_2,a,b)$, with $U_3$ determined. It is invariant under $S$; its fixed subset, obtained by $U_1=U_3$ and $a=b$, has dimension $7$. For $\theta\in C_{\rm mir}\setminus\mathrm{Fix}(S)$, the points $\theta$ and $S\theta$ are distinct regular preimages of the same fingerprint; additional accidental preimages are not excluded by this statement. For $\theta\notin C_{\rm mir}$, the physical channel separates the mirror pair. Thus the reflection sector is completely classified. Any additional collision must be an accidental non-mirror coincidence; it is not generated by a qubit permutation symmetry of the model and is not classified here.

\textbf{Proof.} The code covariance under $S$ is exact at every level. The site-$1$ reference is $\Lambda_1^{\rm Z}(\theta)=\operatorname{Ad}_{U_1e^{-iaZ}}$ and $\Lambda_1^{\rm Z}(S\theta)=\operatorname{Ad}_{U_3e^{-ibZ}}$. Equality of two unitary channels requires $U_3e^{-ibZ}=e^{i\varphi}U_1e^{-iaZ}$ for some phase. Both sides are unimodular: each factor $\exp(-i\sum_j\xi_jP_j)$ with traceless Hermitian generator has determinant one, and $\det e^{-i\theta Z}=1$. Comparing determinants therefore gives $e^{2i\varphi}=1$, so $e^{i\varphi}=\pm1$ and the only alternative to exclude is the sign. A norm bound excludes it: on $\Theta$ each local rotation satisfies $\|U_i-\mathbb I\|\le(|\alpha_i|^2+|\beta_i|^2+|\delta_i|^2)^{1/2}\le0.174$ and the two $ZZ$ factors add at most $0.1$, so both $U_1e^{-iaZ}$ and $U_3e^{-ibZ}$ lie within $r=0.274$ of the identity. Then $\|U_1e^{-iaZ}+U_3e^{-ibZ}\|\ge2-2r=1.45>0$, whereas the sign alternative would make this norm vanish. Hence $\varphi=0$, and no phase freedom survives. Under the resulting condition all five blocks of $\mathbf M_Z$ agree, proving sufficiency. Conversely, equality of $\mathbf M_Z$ implies equality of the site-$1$ reference and hence the displayed condition. The parametrization and dimensions follow because the condition fixes the three real parameters of $U_3$ in terms of $U_1,U_2,a,b$. The fixed subset follows by imposing $S\theta=\theta$. $\square$

\textbf{Proposition 1 (The reflection is the only permutation symmetry of a chain).} Consider the $n$-qubit chain model introduced in Corollary~3 below, $K_\theta=\bigl(\bigotimes_{i=1}^{n}e^{-i(\alpha_iX_i+\beta_iY_i+\delta_iZ_i)}\bigr)\exp\bigl(-i\sum_{i=1}^{n-1}\gamma_iZ_iZ_{i+1}\bigr)$, with all $4n-1$ parameters free, and let $\pi$ be a site permutation for which some parameter map $S_\pi$ satisfies $U_\pi K_\theta U_\pi^\dagger=K_{S_\pi\theta}$ for every $\theta$. Then $\pi$ is the identity or the reversal $R:i\mapsto n+1-i$.

\textbf{Proof.} Conjugation by $U_\pi$ sends $X_i\mapsto X_{\pi(i)}$ and $Z_iZ_{i+1}\mapsto Z_{\pi(i)}Z_{\pi(i+1)}$. Pauli operators of different weight are linearly independent, so the image of the coupling sum must again be a coupling sum of the same family, which requires $\{\pi(i),\pi(i+1)\}$ to be an edge of the chain for every $i$: $\pi$ is an automorphism of the path graph $P_n$. An automorphism maps the two endpoints to the two endpoints, and adjacency then fixes the images of all remaining sites, so there are exactly two possibilities, the identity and $R$. Both occur, because $U_RK_\theta U_R^\dagger=K_{R\theta}$ with $(S_R\theta)$ obtained by reversing the site triples and the couplings. $\square$

For $n=3$ this reproduces the pair $(\mathrm{id},S)$ used above, so the degeneracy is not an accident of three qubits. At $n=2$ the reversal is the site exchange, both references carry the same coupling, and the collision condition of Theorem~2 reduces to $U_1=U_2$, that is to $S\theta=\theta$: the eight-dimensional collision manifold is an $n=3$ feature, not a general one. Any code family and recovery that are themselves covariant under $R$---in particular the repetition codes on a chain with a symmetric recovery---satisfy $\mathcal R_c(R\theta)=\mathcal R_c(\theta)$ at every concatenation depth, so the code-only fingerprint has at least the two preimages $\theta$ and $R\theta$ whenever they differ. The proposition bounds the symmetry-generated sector only; accidental collisions unrelated to a permutation are not excluded at any $n$, and the one-marginal augmentation that closes them for $n=3$ is a candidate rather than a proof at larger $n$.

The reflection sector can nevertheless be closed exactly, and there are two ways to do it. Note first why a unitary single-qubit reference built on $|0\ldots0\rangle$ cannot: with the other qubits held in $|0\rangle$ the reference channel is $\operatorname{Ad}_{U_qe^{-ia_qZ}}$, where $a_q$ is the coupling adjacent to site $q$, so at a collision $U_3e^{-ibZ}=U_1e^{-iaZ}$ the reference channels of qubits $1$ and $3$ coincide identically (numerically to $2\times10^{-16}$). The ambiguity is therefore structural rather than a near-degeneracy, and it is invisible to any single-qubit data built on a $|0\ldots0\rangle$ product reference. Entangling the reference with the chain costs one two-qubit channel (Theorem~3); changing the product state to $|+\rangle$ costs only a preparation change and is the route the selected fingerprint uses (Lemma~1 and Theorem~4).

\textbf{Theorem 3 (One two-qubit marginal removes the reflection ambiguity).} Let $\mathcal R_{12}(\theta)$ be the channel on qubits $(1,2)$ obtained by applying $K_\theta$ to $\rho_{12}\otimes|0\rangle\langle0|_3$ and tracing out qubit $3$, and let $\mathbf M_{\rm aug}=(\mathbf M,\mathcal R_{12})$. Then $\mathbf M_{\rm aug}(\theta)\neq\mathbf M_{\rm aug}(S\theta)$ whenever $S\theta\neq\theta$. Consequently $\mathbf M_{\rm aug}$ has no reflection-generated collision in $\Theta$.

\textbf{Proof.} Suppose $S\theta\neq\theta$ and $\mathbf M_{\rm aug}(\theta)=\mathbf M_{\rm aug}(S\theta)$. Since $\mathbf M$ agrees, Theorem 2 places $\theta$ in $C_{\rm mir}$, so $U_3=U_1e^{-i(a-b)Z}$ and
\[
\begin{aligned}
\mathcal R_{12}(\theta)&=\operatorname{Ad}_{U_1\otimes U_2}
\operatorname{Ad}_{e^{-i(aZ_1Z_2+bZ_2)}},\\
\mathcal R_{12}(S\theta)&=\operatorname{Ad}_{U_3\otimes U_2}
\operatorname{Ad}_{e^{-i(bZ_1Z_2+aZ_2)}} .
\end{aligned}
\]
The coupling layer does not factorize on the two-qubit marginals: with qubit $3$ in $|0\rangle$, $Z_2Z_3$ reduces to $Z_2$, so the $b$ coupling leaves a single-qubit $Z_2$ rotation, and the reflected channel is not obtained by relabeling alone. Using $U_3=U_1e^{-i(a-b)Z_1}$ the second line becomes $\operatorname{Ad}_{U_1\otimes U_2}\operatorname{Ad}_{e^{-i[(a-b)Z_1+bZ_1Z_2+aZ_2]}}$, so equality of the two channels requires the exponentials to agree up to a phase. The exponent difference is $x[Z_1Z_2-Z_1-Z_2]$ with $x=a-b$; its eigenvalues are $-x$ (threefold, for $Z_1=Z_2=1$ and for the two mixed sectors) and $3x$ (once, for $Z_1=Z_2=-1$). The channel ratio therefore has eigenvalues $e^{ix}$ (threefold) and $e^{-3ix}$ (once), and it is a phase times the identity if and only if $e^{4ix}=1$, that is $4x\in2\pi\mathbb Z$ and $x\in(\pi/2)\mathbb Z$. Because $|x|\le0.1<\pi/2$, $x=0$ and hence $a=b$. With $a=b$ the condition $U_3=U_1$ makes the site-$1$ and site-$3$ triples equal and leaves the couplings equal, so $S\theta=\theta$, contradicting the assumption. $\square$

Two remarks. First, the augmented Jacobian keeps rank $11$ at both collision points, because the core submap already has rank $11$ there and adjoining rows cannot lower a rank. Second, because every single-qubit marginal built on the $|0\ldots0\rangle$ reference is provably invariant at the collision ($2\times10^{-16}$), a two-qubit marginal is the cheapest augmentation that keeps the $|0\ldots0\rangle$ reference, although we do not claim minimality outside that class. Its cost is one two-qubit process tomography: $16$ input states times $9$ measurement settings, or $144$ settings executed once per calibration session (the list, the shot allocation and the two-qubit readout correction are given in Sec.~S4 of the Supplemental Material~\cite{SM}). At $1.5\times10^{5}$ shots per setting---the count needed to put the $5.6\times10^{-2}$ separation twenty standard deviations clear---that is $2.2\times10^{7}$ events, three hundred times the $6.8\times10^{4}$ of the seven-channel fingerprint. The factor splits as $144/84=1.7$ from the settings and $185$ from the shots, and the shot factor is not a like-for-like comparison: the marginal is given enough statistics to resolve the very separation that justifies it, whereas the fingerprint needs only $8.1\times10^{2}$ shots per setting for its own $0.01$~rad target. Every setting is a two-qubit circuit, so this is the expensive route rather than the default. Numerically the two-qubit marginals of the collision pair differ by $\max|\Delta\mathcal R_{12}|=5.59\times10^{-2}$; the entries are normalized as $(1/d)\operatorname{Tr}[P_i\mathcal R_{12}(P_j)]$ with $d=4$, exactly as the single-qubit entries with $d=2$, and the maximum runs over all $256$ entries. Against the $2.6\times10^{-3}$ per-component shot noise of its own allocation that is twenty floors, so two-qubit readout calibration, the largest single entry of Table~\ref{tab:nuisances}, has to protect a margin that is not marginal. The remaining unclassified set is therefore restricted to accidental, symmetry-unrelated coincidences of the augmented map, and the reflection sector is solved rather than merely delimited.

The case that can be settled in closed form is a product reference outside the $|0\ldots0\rangle$ class, and it needs no two-qubit channel at all. Prepare the two spectator qubits in \mbox{$|+\rangle$}.

\textbf{Lemma 1 (What a product reference measures).} Let every qubit other than $q$ be prepared in $|+\rangle$, and let $\Lambda_q^{\rm X}(\theta)$ be the resulting reduced channel on site $q$. Then for every $\theta\in\Theta$,
\be
\Lambda_q^{\rm X}(\theta)=\operatorname{Ad}_{U_q}\circ\mathcal D_{s_q},
\ee
with $\mathcal D_s(\rho)=(1-s)\rho+s\,Z\rho Z$, $s_1=\sin^2\gamma_{12}$, $s_3=\sin^2\gamma_{23}$ and $s_2=\sin^2\gamma_{12}+\sin^2\gamma_{23}-2\sin^2\gamma_{12}\sin^2\gamma_{23}$. At a generic point the four-column Jacobian of the $|0\ldots0\rangle$ block with respect to $(\alpha_q,\beta_q,\delta_q,a_q)$ has rank three, the gauge direction $U_q\mapsto U_qe^{-i\tau Z}$, $a_q\mapsto a_q-\tau$ lying in its kernel, so only the product $U_qe^{-ia_qZ}$ is measured; the corresponding Jacobian of the $|+\rangle$ block has rank four and resolves the product into its factors.

\textbf{Proof.} For $q=1$ the spectators are in $|\psi\rangle=e^{-i\gamma_{23}Z_2Z_3}\mbox{$|++\rangle$}=\cos\gamma_{23}\mbox{$|++\rangle$}-i\sin\gamma_{23}\mbox{$|--\rangle$}$, with $\operatorname{Tr}_{23}|\psi\rangle\langle\psi|=\mathbb I$ and $\langle\psi|Z_2|\psi\rangle=0$, because each term carries zero $Z$ magnetization and $\mbox{$|++\rangle$}\perp\mbox{$|--\rangle$}$. Expanding $e^{-i\gamma_{12}Z_1Z_2}=\cos\gamma_{12}\mathbb I-i\sin\gamma_{12}Z_1Z_2$, the cross terms are proportional to $\langle\psi|Z_2|\psi\rangle$ and vanish under the partial trace, leaving $\cos^2\gamma_{12}\rho+\sin^2\gamma_{12}Z\rho Z$; the local unitaries on the traced-out sites drop out of the trace and $U_1$ commutes with it, giving $\operatorname{Ad}_{U_1}\circ\mathcal D_{\sin^2\gamma_{12}}$. Site $3$ is the mirror image, and at site $2$ the two couplings dephase in succession, so $(1-2s_2)=(1-2\sin^2\gamma_{12})(1-2\sin^2\gamma_{23})$. $\square$

\textbf{Theorem 4 (The reflection ambiguity is a degeneracy of the product form).} Let $\theta,S\theta\in\Theta$ with $\theta\neq S\theta$ and $\mathbf M(\theta)=\mathbf M(S\theta)$, so that $\theta\in C_{\rm mir}$ and $U_3=U_1e^{-i(a-b)Z}$ by Theorem~2. Then the single site-$1$ block separates the pair,
\be
\Lambda_1^{\rm X}(\theta)\neq\Lambda_1^{\rm X}(S\theta),
\ee
and so does the site-$3$ block; site $2$ separates no mirror pair, because $s_2$ is symmetric under $a\leftrightarrow b$ and $U_2$ is unchanged by $S$. On $C_{\rm mir}$ the separation $\Delta\Lambda_1=\Lambda_1^{\rm X}(\theta)-\Lambda_1^{\rm X}(S\theta)$ obeys
\be
\max|\Delta\Lambda_1|=\bigl|\sin2(\gamma_{12}-\gamma_{23})\bigr|\bigl(1+O(\theta^2)\bigr).
\ee

\textbf{Proof.} By Lemma~1, $\Lambda_1^{\rm X}(\theta)=\operatorname{Ad}_{U_1}\mathcal D_{s_a}$ and $\Lambda_1^{\rm X}(S\theta)=\operatorname{Ad}_{U_3}\mathcal D_{s_b}$ with $s_a=\sin^2a$, $s_b=\sin^2b$, $a=\gamma_{12}$ and $b=\gamma_{23}$. On $C_{\rm mir}$, $U_3=U_1e^{-icZ}$ with $c=a-b$, and $[\mathcal D_s,\operatorname{Ad}_Z]=0$, so equality of the two channels would force
\be
\mathcal D_{s_a}=\operatorname{Ad}_{e^{-icZ}}\mathcal D_{s_b}=\mathcal D_{s_b}\operatorname{Ad}_{e^{-icZ}}.
\ee
The PTM of $\mathcal D_s$ is diagonal with entries $1,1-2s,1-2s,1$ in the order $(I,X,Y,Z)$, and $\operatorname{Ad}_{e^{-icZ}}$ is orthogonal, so the singular values of the right-hand side are $1,|1-2s_b|,|1-2s_b|,1$. Equality of channels therefore requires $s_a=s_b$, since $s\le\sin^20.05<1/2$ on $\Theta$. With $s_a=s_b$ and $\mathcal D_{s_a}$ invertible the displayed relation reduces to $\operatorname{Ad}_{e^{-icZ}}=\mathrm{id}$, that is $c\in\pi\mathbb Z$; but $|c|=|a-b|\le0.1<\pi$, so $c=0$, hence $a=b$ and $U_3=U_1$, which gives $S\theta=\theta$ and contradicts the assumption. For the leading term, the two PTMs differ by $\mathcal R(U_1)[\mathcal D_{s_a}-\mathcal D_{s_b}\mathcal Z_{2c}]$, where $\mathcal Z_{2c}$ is the $Z$ rotation by $2c$ acting on the $(X,Y)$ block: its $Y$ entry is $-\sin2c\,(1-2s_b)$, while its diagonal entries differ only at $O(\theta^2)$, since $s_b-s_a=(b-a)(a+b)+O(\theta^4)$. $\square$

\textbf{Corollary 4 (The selected fingerprint has no reflection collision).} Let $\mathbf M$ be the seven-channel map above. Then $\mathbf M(\theta)\neq\mathbf M(S\theta)$ for every $\theta\in\Theta$ with $S\theta\neq\theta$, and $\operatorname{rank}J(\theta)=11$ on $\Theta\setminus V$. Consequently the inversion uses a single point on the mirror orbit and no appended channel: the reflection ambiguity is a property of the $|0\ldots0\rangle$ frame, not of the family. \emph{Proof.} If $\theta\notin C_{\rm mir}$ the site-$1$ reference already separates the pair (Theorem~2); if $\theta\in C_{\rm mir}$ then $\Lambda_1^{\rm X}$ separates it (Theorem~4). Rank is Theorem~1. $\square$

\textbf{Remark (Cost of the $|+\rangle$ reference).} The site-$q$ block is one single-qubit process tomography---twelve settings, against the $144$ of $\mathcal R_{12}$---and needs no two-qubit circuit; only the preparation of the two spectator qubits changes, and no additional gate is applied to the block. The proof also shows why the change must be of this kind: the interference term that would restore a unitary component is proportional to the $Z$ magnetization of the spectator state, which vanishes for \mbox{$|+\rangle$} and not for a generic product state. The price is a state-preparation budget rather than a gate budget. With $\varepsilon_H$ the amplitude error of the two spectator $\pi/2$ pulses, the spectator magnetization is $\langle Z\rangle\approx2\varepsilon_H$, so the unitary component leaked back into the reference is of order $2\varepsilon_H\sin\gamma\sim6\times10^{-5}$ for $\varepsilon_H=10^{-3}$, two orders of magnitude below the $5.6\times10^{-2}$ separation that Theorem~4 establishes; any $\varepsilon_H\ll0.1$ is therefore adequate, and the $|+\rangle$ reference is preparation-limited only in a regime where the $|0\rangle$ reference is already unusable.

\textbf{Remark (What the two frames each miss).} With the other qubits in $|0\ldots0\rangle$ the reference takes the compact form $\Lambda_q^{\rm Z}=\operatorname{Ad}_{V_q}$ with $V_q=U_qe^{-is_qZ}$, $s_1=\gamma_{12}$, $s_2=\gamma_{12}+\gamma_{23}$ and $s_3=\gamma_{23}$, so the three-site datum, $16$ numbers per site, depends on $\theta$ only through the three products $V_q$. It is therefore exactly invariant under $\gamma_{12}\mapsto\gamma_{12}+\Delta$, $\gamma_{23}\mapsto\gamma_{23}+\Delta'$ and $U_q\mapsto U_qe^{i\Delta_qZ}$ with $(\Delta_1,\Delta_2,\Delta_3)=(\Delta,\Delta+\Delta',\Delta')$, a two-parameter family verified at the $10^{-16}$ level: the $|0\ldots0\rangle$ frame has rank $3n=9$ with this family as its kernel, and neither bond coupling is visible in it on its own. The $|+\rangle$ datum of Lemma~1 is complementary in both directions: it resolves the products into factors, because $\operatorname{Ad}_W\mathcal D_s=\mathcal D_{s'}$ forces $s'=s$ and $W=\pm\mathbb I$, and it is exactly even in each coupling, because only $\sin^2\gamma$ enters, so it fixes $|\gamma_{12}|$ and $|\gamma_{23}|$ and not their signs. The two frames are therefore not interchangeable. A single $|+\rangle$ reference kills one of the two kernel directions, and the two at sites $1$ and $3$ kill both: $\Delta$ is fixed by site $1$ and $\Delta'$ by site $3$, so $\Lambda_2^{\rm X}$ is redundant, which is why the selected fingerprint carries five references and not six. The five together have rank $11$ with no kernel and no discrete ambiguity, and invert $\theta$ in two steps: site-$1$ and site-$3$ of the $|+\rangle$ frame give $U_1$, $U_3$ and $|\gamma|$; the signs of the couplings then follow from the $|0\ldots0\rangle$ frames, which see them linearly; and $U_2$ follows from $V_2$ once $\gamma_{12}+\gamma_{23}$ is known. Adding the two code channels raises $s_{\min}$ from $1.06$ to $1.76$ without changing the rank, so their contribution is statistical and diagnostic rather than identifying.

\textbf{Status of global uniqueness.} The complementary response of the two frames is useful for designing probes and for interpreting which parameter directions are weakly constrained, but it is not a global injectivity proof: a perturbative statement that a direction hidden by one frame appears at leading order in the other does not control nonperturbative cancellations. The rigorous statement therefore remains local: on $\Theta\setminus V$, $\mathbf M$ is an immersion and has a local inverse. The reflection sector is closed by Corollary~4 rather than left as an open problem, so what remains is the possibility of accidental collisions between points unrelated by any symmetry. Random multistart inversion and the $500$ Jacobian samples are diagnostics for such collisions but cannot exclude them, and we do not claim they are absent. We therefore make no conjecture of unrestricted global injectivity, and we do not use the finite-fiber language: determining all global preimages, or excluding accidental coincidences outside the reflection sector, remains open.

\textbf{Remark (Higher-order visibility and conditioning).} Directions corrected by a code enter the logical fingerprint only at higher order, so the code-only fingerprint is ill-conditioned: over the same $500$ samples $s_{\min}$ runs from $4.6\times10^{-4}$ to $3.4\times10^{-2}$ (median $8.8\times10^{-3}$) and the condition number from $1.4\times10^{2}$ to $1.1\times10^{4}$ (median $5.5\times10^{2}$); the product references supply first-order responses to the same directions and bring the $112\times11$ Jacobian to a condition number between $3.6$ and $3.9$ over the whole of $\Theta$ in our samples. Protection and parameter sensitivity are complementary features of the same projection, and the complete data set restores a direction-dependent but experimentally usable precision.
For every $\theta\in\Theta\setminus V$, there is an open neighborhood $U$ of $\theta$ on which $\mathbf M$ is injective and has a continuous local inverse. Thus exact fingerprint data determine the noise parameters uniquely within a sufficiently small neighborhood of any regular point. This does not assert that the preimage is globally unique or finite: the compensated collision in Appendix C provides two regular points with the same fingerprint. Under finite single-qubit QPT statistics, a least-squares or maximum-likelihood inversion initialized inside the local basin yields a locally consistent estimate, with precision governed by the local Jacobian and the Fisher information discussed below.

\textbf{Corollary 2 (Constant cost per fingerprint for a fixed model).}
Each fingerprint is a single-qubit channel and therefore requires a fixed, model-independent number of QPT settings, and the total number of QPT experiments is constant for a fixed model. The rank count does not depend on how many channels are read beyond eleven: the five product references already attain rank $11$, and the two code channels, which are the object of the measurement, raise $s_{\min}$ from $1.06$ to $1.76$ without adding an identifying equation. Reading more channels is therefore a statistical and diagnostic choice rather than an identifiability requirement, which is what makes the selection of seven a design decision with a measurable answer (Appendix~D). This statement does not imply constant cost for a growing model: the number of required fingerprints is set by the parameter dimension.

\textbf{Corollary 3 (Multi-qubit extension and cost accounting).}
We now state the cost accounting for an $n$-qubit repetition-code block. This is not a general scalability theorem. The polarization tendency is assumed for the same small-angle unitary family, now with stabilizers generated by $Z_i Z_{i+1}$ (bit-flip type) or $X_i X_{i+1}$ (phase-flip type) on a chain. Throughout this section, $n$ denotes the number of physical qubits in a single repetition-code block, and $\ell$ denotes a concatenation level. The noise model to be recovered is defined on the $n$ qubits of the innermost block:
\[
\begin{aligned}
K_\theta={}&\left(\bigotimes_{i=1}^{n}
e^{-i(\alpha_i X_i+\beta_i Y_i+\delta_i Z_i)}\right)\\
&\times\exp\!\left(-i\sum_{i=1}^{n-1}\gamma_i Z_i Z_{i+1}\right),
\end{aligned}
\]
with $\theta=(\alpha_i,\beta_i,\delta_i,\gamma_i)\in\Theta_n$. The parameter domain $\Theta_n$ is compact and its dimension depends only on $n$, never on the total device size. A larger processor is viewed as a collection of $B$ independent $n$-qubit blocks. All blocks share the same parameters $\theta$; this uniformity assumption makes the parameter set $\Theta_n$ independent of $B$. Under this assumption, repeated measurements on different blocks reduce statistical uncertainty but do not enlarge the parameter space.

This uniform-calibration mode should be distinguished from a pair-resolved diagnostic mode. On a large chip, the number of possible stray two-qubit couplings can grow combinatorially, up to $O(n^2)$ without connectivity restrictions. A diagnostic protocol would instead assign parameters to selected blocks or coupling hypotheses and use the directional polarization signatures to localize and classify them. Such physical attribution is not available from a single global performance metric. The constant-overhead statement proved here relies on shared parameters and does not automatically extend to arbitrary pair-resolved tomography; the number of probes required in that diagnostic regime must be analyzed separately.

Every code used in this protocol encodes exactly one logical qubit. Thus, one measurement-free QEC round on an $n$-qubit block produces exactly one effective channel $\mathcal R_{\mathrm{eff}}(\theta)$, never a collection of $n$ channels. To identify the parameters in $\Theta_n$, we execute $L$ distinct QEC configurations on the same block, using different code orientations, stabilizer orderings, or concatenation depths. Each configuration yields one effective channel $\mathcal R_{\mathrm{eff}}^{(j)}(\theta)$, and the fingerprint is
$$m(\theta)=\left(\mathcal R_{\mathrm{eff}}^{(1)}(\theta),\ldots,
\mathcal R_{\mathrm{eff}}^{(L)}(\theta)\right).$$
Each effective channel is reconstructed by single-logical-qubit QPT, whose number of measurement settings is constant. Therefore, the tomographic cost per fingerprint is constant. The number of fingerprints $L$ is constrained by the dimension of $\Theta_n$: a general single-qubit trace-preserving channel has at most 12 real degrees of freedom, so local identification of $4n-1$ parameters requires at least
\[
12L\ge 4n-1 .
\]
This is a necessary dimension count, not a sufficient construction. For the present $n=3$ model, seven fingerprints are ample; for a general block, $L$ grows at least linearly with $n$. The number of fingerprints does not grow with the number $B$ of identical blocks when all blocks share one fixed parameter vector, but it does grow if each block or coupling is independently parameterized. Thus only the cost per fingerprint, not the total model-identification cost, is constant.

The key local requirement is that the Jacobian $J_\theta=\partial m/\partial\theta$ of the stacked fingerprint map $$m:\Theta_n\to\mathbb{R}^{D}$$ have full column rank on an open dense subset of $\Theta_n$. Complementary polarization provides the physical design principle: each additional code orientation should expose directions that the existing fingerprints suppress. Theorem 1 certifies this rank condition for the present three-qubit, 11-parameter model; for a general $n$-qubit model it must be certified separately. Global injectivity is false in general, as the exact collision shows, and is not implied by local rank. The construction invokes no compressed-sensing sparsity or low-rank reconstruction. It does, however, rely on the explicit unitary family $K_\theta$; in particular, its Choi matrix is rank one by construction. That model restriction must be stated rather than hidden behind a claim of assumption-free tomography.

The dominant error source is gate calibration. Let $M_l$ be the total gate count of one measurement-free encoding-correction round on an $n$-qubit block at concatenation level $l$. For repetition-type codes, $M_l=c_l n$ with $c_l$ a level-dependent constant. If every physical gate has calibration uncertainty $\delta\eta_{\mathrm{cal}}$, the newly introduced effective-channel error at level $l$ is $s_l M_l^{p}\,\delta\eta_{\mathrm{cal}}$, where $s_l$ is the gate-error sensitivity and $p=\tfrac12$ for stochastic calibration errors, $p=1$ for worst-case coherent accumulation.
The accuracy of a concatenated configuration must be computed recursively, because the residual error of the previous level enters the next level as an input. Let $\varepsilon_l$ be the effective-channel error after level $l$, and $\kappa_l$ the amplification with which level $l$ propagates the incoming channel error from level $l-1$. Then
$$\varepsilon_0=0,\qquad
\varepsilon_l=\kappa_l\,\varepsilon_{l-1}+s_l M_l^{p}\,\delta\eta_{\mathrm{cal}}.$$
Under the uniform approximation $\kappa_l\approx\kappa$, $s_l\approx s$, $M_l\approx M$,
$$\varepsilon_l\approx\frac{1-\kappa^{l}}{1-\kappa}\,s M^{p}\,\delta\eta_{\mathrm{cal}}.$$
The single-level formula corresponds to $l=1$, which is the design used here. A two-level implementation would add two distinct contributions---the deeper gates, scaling as $M_2^{p}\,\delta\eta_{\mathrm{cal}}$, and the first-level residual propagated as $\kappa_2\,\varepsilon_1$---so different concatenation depths have genuinely different precision, and it is the deepest level that sets the budget in that case.
If $\kappa<1$, the concatenated error saturates within a factor $1/(1-\kappa)$ of the single-level value. If $\kappa=1$, the error accumulates linearly in $\ell$. Concatenation does not improve parameter identifiability: later channels are deterministic functions of the first-level ones. Its only possible benefit is additional independent statistical information, which must be compared with the added gates, calibration burden, and accumulated error. The recursion above determines, for each measured configuration, the calibration precision required to keep its fingerprint error below a stated statistical target.

For the present fingerprint, this recursion gives an illustrative experimental benchmark at the representative parameter point used below. The shot budget of the measurement is $N=1.5\times10^{3}$ per component, so $\sigma_{\rm shot}=2.540\times10^{-2}$. With $M_0=1$ for the five product references and $M_1=6$ for the two code channels, unit sensitivities and $\delta=\delta\eta_{\rm cal}$, the level-0 error is $e_0=\delta$ and the code-channel error is $e_1=2.4495\delta$ for stochastic or $6\delta$ for coherent accumulation; the second level no longer appears, and this is the quantitative form of the difference from the concatenated design, whose $18$ gate uses per repetition made the deepest level the calibration bottleneck. Averaging over the seven channels at the PTM-component RMS level gives
\[
\begin{aligned}
b_p&=\sqrt{\frac{5e_0^2+2e_1^2}{7}}=C_p\delta,\\
C_{1/2}&=1.558,\qquad C_1=3.317 .
\end{aligned}
\]
Thus $b_p\leq\sigma_{\rm shot}$ allows $\delta\leq1.63\times10^{-2}$ for stochastic accumulation or $\delta\leq7.66\times10^{-3}$ for coherent accumulation. A twofold amplitude margin halves both. A per-gate residual of $10^{-3}$, which is the optimistic end of the range reported for optimized devices, then sits more than an order of magnitude below the shot noise rather than at the bottleneck, and the same holds at the per-channel level: considered alone, the code channels stay below $\sigma_{\rm shot}$ for $\delta\leq1.04\times10^{-2}$ (stochastic) or $4.23\times10^{-3}$ (coherent), and the references for $\delta\leq2.54\times10^{-2}$. This is the quantitative form of the claim that the probe is shot-noise limited. The thresholds constrain the RMS fingerprint error relative to shot noise; they do not guarantee $0.01$~rad parameter accuracy for an adversarially aligned coherent bias. For that, the conservative triangle bound $4(5e_0+2e_1)=68\delta$, inverted through $s_{\min}=1.7554$, gives the much stricter sufficient condition $\delta\leq2.58\times10^{-4}$ for $\|\Delta\theta\|_2\leq0.01$~rad, and $\delta\leq1.29\times10^{-3}$ for $0.05$~rad. The difference between the two readings is the point of the appendix: the \emph{RMS} statement answers whether the fingerprint is above its noise, the triangle bound answers whether a bias aligned with the weakest direction can move a parameter, and only the second is a requirement on calibration. These numbers are illustrative scaling estimates rather than platform specifications. Actual gate counts, sensitivities, inter-level correlations, drift, and state-preparation and measurement errors must be measured and inserted into the recursion. The full budget, including arbitrary-direction coherent bias and nonuniform one- and two-qubit gate errors, is given in Appendix A.

The construction needs only nearest-neighbor two-qubit gates: on a chain layout every stabilizer spans adjacent qubits and measurement-free recovery uses transversal CNOTs with controlled Pauli operations on logical ancillas, so no nonlocal multi-controlled-gate decomposition is required. This does not establish constant total cost for a growing device, since $4n-1$, the number of fingerprints, and the inversion dimension all grow.

\begin{figure}[!tb]
\centering
\includegraphics[width=0.47\textwidth]{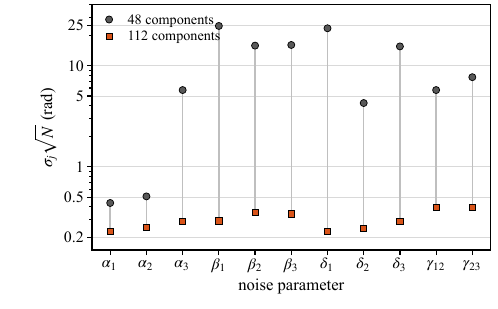}
\caption{Per-parameter Cram\'er--Rao coefficient $\sigma_j\sqrt{N}$ at the reference point $\theta_{\rm res}$ for the $48$-component core (grey circles) and the selected $112$-component fingerprint (orange squares); note the logarithmic scale. Smaller is better. Every coefficient of the selected map lies between $0.23$ and $0.40$, whereas the core spans two orders of magnitude, so the additional references flatten the anisotropy rather than only shifting it. The value quoted in the text is the worst one, $\sigma\sqrt{N}\le0.394$, which is attained by the two couplings.}
\label{fig:crb}
\end{figure}

\emph{Numerical verification.}
We corroborate the identifiability guarantee by a synthetic numerical implementation of the three-qubit protocol. The noise is the unitary channel with parameter domain $\Theta=[-0.1,0.1]^{9}\times[-0.05,0.05]^{2}$. The observation map $\mathbf M(\theta)$ stacks the seven real PTMs $\Lambda_1^{\rm Z},\Lambda_2^{\rm Z},\Lambda_3^{\rm Z},\mathcal R_X,\mathcal R_Z,\Lambda_1^{\rm X},\Lambda_3^{\rm X}$ in the fixed order of Appendix C, giving a real fingerprint vector of dimension $D=112$ for the $11$-dimensional parameter vector $\theta$.

At each of $500$ randomly sampled parameter points $\theta_{i}\in\Theta$, we construct the Jacobian
\[
J(\theta_{i})=\frac{\partial \mathbf M}{\partial\theta}\Big|_{\theta_{i}}\in\mathbb{R}^{112\times11}
\]
by central differences with step $\varepsilon=10^{-5}$:
\[
\partial_{j}\mathbf M(\theta)\simeq\frac{\mathbf M(\theta+\varepsilon e_{j})-\mathbf M(\theta-\varepsilon e_{j})}{2\varepsilon}.
\]
The linearization error is estimated by comparing the exact fingerprint difference with the Jacobian prediction for a random perturbation $\xi$ with $|\xi|\leq10^{-7}$; the relative error
\[
\frac{\lVert \mathbf M(\theta_{i}+\xi)-\mathbf M(\theta_{i})-J(\theta_{i})\xi\rVert}
{\lVert \mathbf M(\theta_{i}+\xi)-\mathbf M(\theta_{i})\rVert}
\]
is of order $10^{-7}$ in all tested cases, confirming that the numerical Jacobian accurately captures the local response of the fingerprint.

The results are uniform across the entire sample:
\begin{itemize}
\item $\mathrm{rank}\,J(\theta_{i})=11$ for all $500$ points, i.e., full column rank at every sample;
\item $\ker J(\theta_{i})=\{0\}$, so no first-order perturbation of the $11$ parameters is invisible to the fingerprint;
\item the smallest singular value $s_{\min}(\theta_{i})$ is strictly positive at all samples, ranging from $1.748$ to $1.828$, with a median of $1.763$.
\end{itemize}

Across the $500$ samples, $s_{\min}$ stays between $1.748$ and $1.828$---a spread of $4.6\%$---and the condition number between $3.63$ and $3.85$, so the conditioning is not merely bounded away from rank deficiency but uniform over $\Theta$. Figure~\ref{fig:crb} instead shows where that conditioning is spent: the per-parameter Cram\'er--Rao coefficients separate the cheap directions from the expensive ones far more informatively than a histogram of $s_{\min}$. All Fisher, calibration, and resource numbers in this paper refer to one resource-comparison point, drawn independently and uniformly per coordinate from $\Theta$,
\[
\begin{aligned}
\theta_{\rm res}=(&-0.0633,\;0.0807,\;0.0505,\;0.0437,\\
&0.0191,\;0.0767,\;-0.0613,\;0.0940,\\
&-0.0602,\;0.0240,\;0.0366);
\end{aligned}
\]
the decoder and inversion tests use a separate decoder point $\theta_{\rm dec}$, specified in Appendix~B. At $\theta_{\rm res}$ the singular values are \[
\left\{
\begin{aligned}
&6.701,\;6.100,\;4.572,\;4.455\\
&4.054,\;3.995,\;3.992,\;3.175\\
&2.826,\;2.600,\;1.755
\end{aligned}
\right\},
\]
with a condition number $3.82$, and the conditioning is uniform rather than merely finite: over the $500$ sampled Jacobians the smallest singular value stays in $[1.748,1.828]$ and the condition number in $[3.63,3.85]$, a spread of $4.6\%$ in $s_{\min}$. At the origin the Jacobian is still of full rank, with $s_{\min}=1.748$; it is the $48$-row core that degenerates there, to rank $5$, so the origin is singular for the core and regular for the full map. The three families are deficient in three different ways, and it is worth separating them because only the union is complete. The two code channels alone have full rank $11$ and $s_{\min}=1.5\times10^{-2}$, yet they cannot separate $\theta$ from $S\theta$: they are exactly invariant under the reflection (Theorem~2), so full rank and local injectivity coexist with a twofold global fiber. The three $|0\ldots0\rangle$ references alone have rank $9$, with the two-parameter kernel of the remark after Theorem~4, and the two $|+\rangle$ references alone have rank $8$; neither can stand in for the other. The five references together have rank $11$ and $s_{\min}=1.06$, no kernel and---by Corollary~4---no mirror collision, so the reflection ambiguity is not removed from the seven-channel datum by an appended channel but by the frame that is already in it; adding the code channels then raises $s_{\min}$ to $1.76$.

Because the three-channel core is the cheaper fallback, its own conditioning matters as well. At the same point $J_{48}$ has $s_{\min}=2.79\times10^{-2}$ and condition number $1.9\times10^{2}$, so the four additional references buy a factor $63$ in $s_{\min}$ and a factor $50$ in condition number, at four extra gate uses per repetition and $48$ extra settings. The per-parameter coefficients at $\theta_{\rm res}$ lie between $0.229$ and $0.394$ for the selected map, against $0.44$ to $24.6$ for the core (Fig.~\ref{fig:crb}): the design has flattened the anisotropy rather than only shifting it, and the weakly constrained quantities are now the two couplings themselves ($0.394$ each, on the weakest singular direction $0.60(\gamma_{23}-\gamma_{12})+0.37(\delta_{1}-\delta_{3})$) rather than the $X$ and $Z$ angles of the outer qubits. A bias aligned along that direction is amplified by $1/s_{\min}$, which is the direction to which the conservative calibration bound of Appendix A refers. These observations are consistent with the generic local-identifiability claim of \textbf{Theorem 1} and with $V$ being of measure zero. They do not establish global uniqueness; indeed the exact collision in Appendix C lies outside $V$ and refutes that stronger statement for the $|0\ldots0\rangle$ frame, which is why the design carries two frames.

\emph{Parameter recovery by closed-loop inversion.}
We tested end-to-end recovery from both exact and noisy fingerprints. For two randomly drawn generic ground-truth vectors $\theta^{\ast}\in\Theta$, Gauss--Newton inversion of $\mathbf M(\theta)=\mathbf M(\theta^{\ast})$ with a backtracking line search converges to $\theta^{\ast}$ with final residuals $\sim10^{-15}$ from starts in its basin. Starts placed near the mirror image $S\theta^{\ast}$ do not converge to a zero-residual solution at these generic points (final residuals $2.4\times10^{-3}$ and $7.4\times10^{-3}$), because the references differ there. At a point of $C_{\rm mir}$ the codes and the $|0\ldots0\rangle$ references agree with those of $S\theta$ and only the $|+\rangle$ references separate the pair, by $5.6\times10^{-2}$ at the certified example of Appendix C, and that is the margin by which the inversion tells the two basins apart. In units of the measurement, the two $|+\rangle$ blocks together move by $\|\Delta\|=1.1\times10^{-1}$ between the colliding points, which is $4.4$ standard deviations of a known direction at the recommended $N=1.5\times10^{3}$ shots per setting ($\sigma_{\rm shot}=2.54\times10^{-2}$), and the margin grows as $\sqrt N$. A residual near $10^{-15}$ proves that a preimage was found, never that it is unique. Failed starts and local minima were retained in the inversion log rather than counted as successes.

\emph{Search for accidental collisions.} Because the reflection family is resolved by Corollary~4, the only remaining ambiguity would be an accidental coincidence between points unrelated by symmetry. We searched for one directly: for $10$ random targets in $\Theta$ we started $100$ Gauss--Newton searches each from random points and retained only solutions inside $\Theta$ with residual below $10^{-8}$. All $1000$ searches converged to the target itself, and no second in-domain solution appeared. That is a stronger outcome than the concatenated design gave, where the same driver recovered the target for $37$ of $40$ targets and left three to basins of attraction; the difference is the conditioning, not the algorithm. Two cheap strengthenings bound what remains. (i) Each independent PTM component is a Laurent polynomial in the eleven generator phases with per-variable exponent span at most two (each generator enters the channel unitary once and its adjoint once); after the standard $t=\tan(\theta/2)$ rationalization and clearing denominators, a generic set of eleven defining equations has per-variable degree at most two and total degree at most twenty-two, so B\'ezout's theorem, applied to eleven equations of those total degrees, bounds the generic complex fiber cardinality by the product of the eleven degrees, $(22)^{11}$---loose, but finite and explicit. (ii) The same driver can be repeated at the five sampled Jacobians with the worst condition numbers without new code, and the archived log records the seeds, so the run is a matter of compute rather than of design. A search of this size cannot prove injectivity, and $37/40$ in the earlier design shows that a hit rate has to be quoted alongside a negative result; what the present run does is convert the surviving hypothesis into a quantitative statement rather than an absence of evidence.

For the shot-noise analysis we use a deliberately simple surrogate: each PTM component in the 112-vector has independent Gaussian error $\sigma=1/\sqrt{N}$. Here $N$ is the effective number of shots assigned to each component after all QPT settings and channels are combined; it is not the total physical shot count. A measurement-level analysis must instead use the classical Fisher information
\[
I_{ij}(\theta)=\sum_{s}\sum_{o}\frac{N_s}{p(o|s;\theta)}
\frac{\partial p(o|s;\theta)}{\partial\theta_i}
\frac{\partial p(o|s;\theta)}{\partial\theta_j},
\]
with explicitly chosen input states, measurement settings, and shot allocation. Trace preservation and the channel constraints also reduce the number of independent components. The Gaussian surrogate is therefore a local numerical model, not a complete measurement protocol.

At $\theta_{\rm res}$, the surrogate CRB $\mathrm{Cov}(\hat\theta)\succeq\sigma^{2}(J^{T}J)^{-1}$ gives per-parameter coefficients $\sigma_j\sqrt{N}$ of
\[
\begin{aligned}
&0.2291,\;0.2910,\;0.2290,\;0.2494,\;0.3533,\;0.2476,\\
&0.2886,\;0.3410,\;0.2883,\;0.3936,\;0.3935
\end{aligned}
\]
in the order $(\alpha_1,\alpha_2,\alpha_3,\beta_1,\beta_2,\beta_3,\delta_1,\delta_2,\delta_3,\gamma_{12},\gamma_{23})$. Reaching $0.01$~rad on every parameter at this point therefore requires $N\simeq1.5\times10^{3}$ shots per component---the bottleneck is now either coupling, at $0.394$, not a single-qubit angle---and reaching $0.05$~rad requires only $62$. At twelve settings per channel and seven channels that is $84N=1.3\times10^{5}$ events. That budget is statistical. A non-unitary component of probability $p$ saturates the fit at $N_{\rm sat}=1/\mu^{2}\approx5.7\times10^{4}$ for $p=10^{-3}$, which is now \emph{above} the shot budget rather than forty to a hundred times below it: at the operating point the model error is no longer the binding constraint, and the ordering has reversed with respect to the concatenated design. To calibrate what the target means: $0.01$~rad is $10\%$ of the half-width of $\Theta$ and about $25\%$ of a typical parameter value such as $\alpha\sim0.04$~rad. It is therefore a coarse but non-trivial requirement, sufficient to fix the sign and rough magnitude of every term, and the numbers below should be read that way rather than as high-precision metrology. This is an operating-point statement, not a domain-wide guarantee. Because the conditioning is uniform, the sample does not degrade it: at the worst of the $500$ Jacobians the same target needs $N=1.57\times10^{3}$, $1.4\%$ more than the $1.55\times10^{3}$ at $\theta_{\rm res}$ (the singular-value spread alone would predict $0.9\%$, the budget being set by the weakest parameter direction rather than by $s_{\min}$), and the necessary bound $N\gtrsim1/[11(0.01)^{2}s_{\min}^{2}]$ that dominated the concatenated design at $7.2\times10^{7}$ falls to $3.0\times10^{2}$.

The surrogate can be replaced by a measurement-level calculation at modest cost. For the actual twelve-setting protocol, namely the four input states $|0\rangle,|1\rangle,|+\rangle,|+\mathrm{i}\rangle$ times the three measurement bases, the multinomial Fisher information above is evaluated by finite differences of the outcome probabilities. With ideal state preparation and readout, the resulting CRB is a factor $0.72$ of the surrogate value for the seven-channel set and $0.32$ for the three-channel core, in both cases as the median over parameters; the surrogate is therefore conservative, and the shot counts quoted here are upper estimates rather than optimistic ones. At the same target the measurement-level requirement is $8.1\times10^{2}$ shots per setting for the seven-channel set, that is $12\times7\times8.1\times10^{2}\approx6.8\times10^{4}$ physical events, to be compared with the $16\times7\times1.5\times10^{3}\approx1.7\times10^{5}$ nominal events implied by reading the per-component convention literally. The convention still matters---a factor $2.5$ here, against $17$ for the concatenated design, in the opposite direction---so a statement in effective shots per fingerprint remains ambiguous unless the convention is fixed. We retain the surrogate everywhere else because it treats all configurations identically, and because state-preparation and readout errors, excluded from the measurement-level calculation, act in the opposite direction. Conditioning controls the amplification of both statistical and calibration errors, so recovery is not limited by shot statistics ``rather than'' conditioning. At the representative point, independent random calibration residuals of $10^{-4}$ per gate produce a maximum per-parameter calibration RMS of $6.1\times10^{-5}$~rad, three orders of magnitude below the $0.01$~rad allowance; coherent accumulation gives $1.3\times10^{-4}$~rad, still seventyfold below. The conservative worst-direction sufficient condition is $2.58\times10^{-4}$ per gate, the same statement as $\mu<s_{\min}\varepsilon=1.76\times10^{-2}$ with $\varepsilon=0.01$~rad. A stochastic component at $p=10^{-3}$ has $\mu=4.19\times10^{-3}$, which now \emph{satisfies} that allowance by a factor $4.2$ rather than exceeding it; the allowance is reached only at $p=4.2\times10^{-3}$, and for a dephasing component at $p=7.2\times10^{-3}$. The operationally meaningful form remains the per-source pair $|\delta\theta|<\varepsilon$ with $N<N_{\rm vis}=112/r^{2}$. For the readout term the same accounting fixes an implicit SPAM threshold: a $2\%$ readout error shifts the generator by $2.9$~mrad, so the $0.01$~rad target requires readout below $\approx7\%$, two orders of magnitude looser than the concatenated design needed, because the reference entries are first-order responses of order unity rather than third-order residues. Increasing $N$ reduces only the statistical contribution and does not remove a fixed calibration bias. The per-parameter scatter quoted above is therefore a shot-noise scatter: state preparation and readout enter through this threshold and through the readout row of Table~\ref{tab:nuisances}, not as an additional Fisher term.

\emph{Drift.} The seven channels are acquired sequentially, so a slowly drifting parameter set makes the blocks mutually inconsistent, and unlike calibration bias this mismatch cannot be reduced by more shots. Linearizing, a drift that shifts block $c$ by $\delta\theta_c$ perturbs the reconstructed fingerprint by $J_c\delta\theta_c$ and biases the estimate by $(J^TJ)^{-1}\sum_cJ_c^TJ_c\,\delta\theta_c$. For a constant rate $r$ along a fixed direction and a gate-weighted sequential schedule the bias is $rT\,Ke$ with $\|K\|_2=1.29$, so holding the bias below one third of the target requires $rT\le2.6\times10^{-3}$~rad. For a one-hour acquisition this is $2.6$~mrad per hour, an order of magnitude above the roughly $10^{-4}$~rad per hour typical of stabilized control fields for the angles considered here, and the tolerance scales inversely with total measurement time. The near-orthogonality of the seven blocks is what buys this: in the concatenated design the four second-level blocks carried almost no first-order response, so a drift that moved them was amplified through the inverse of a badly conditioned normal matrix and $\|K\|_2$ was $10.3$. Interleaving the repetitions of the seven channels instead of blocking them makes every channel sample the same average time and brings $\|K\|_2$ to $0.5$, a factor $2.6$. Randomizing the order on top of interleaving turns any residual linear drift into zero-mean noise rather than a systematic offset, at the cost of more frequent reconfiguration; and zero-mean jitter then enters only at second order, $1$~mrad of Gaussian drift in a coupling moving the fitted generator by $1.1\times10^{-7}$~rad ($\mu=5.7\times10^{-6}$), far below shot noise at any realistic budget. Drift should therefore be handled as a scheduling constraint rather than as an irreducible systematic floor.

\emph{Model space versus generator space.}---The eleven parameters are eleven of the sixty-three nontrivial Pauli directions of a three-qubit generator, so the family is a subspace rather than a basis, and the operative question is what the fit reports when the true noise contains a component outside it. We answer by construction: a term $\exp(-\mathrm{i}aG)$ is added on top of the model for each of the remaining fifty-two directions, the fingerprint is recomputed, and the discrepancy $\delta\mathbf M$ is split into the part the model absorbs, $\delta\theta=J^{+}\delta\mathbf M$, and the part it cannot, $\rho=\|(I-JJ^{+})\delta\mathbf M\|/\|\delta\mathbf M\|$. Both are first order and no optimizer enters. With the selected fingerprint the amplification $\max|\delta\theta|/a$ has median $0.42$ and \emph{maximum} $0.97$: no direction of the fifty-two is amplified above its own size, where the concatenated design amplified thirty-seven of them, by up to a factor $65$. Nothing is silent either. The smallest $\rho$ at $\theta_{\rm res}$ is $0.11$ ($Y_{2}Z_{3}$), and over the fifteen sampled points of Appendix~F only two directions, $Y_{2}Z_{3}$ and $Z_{1}Y_{2}$, fall below $0.1$ at any point, at one and two of them respectively and with a minimum of $0.053$; their amplification stays below unity there, so what remains is a bias of the order of the term itself rather than a deleted response. The experimentally common $Z_{1}Z_{3}$ crosstalk illustrates the change: $1$~mrad is now reported as $\delta_{2}=-0.80$, $\gamma_{12}=+0.51$, $\gamma_{23}=+0.47$, $\delta_{3}=+0.33$ and $\delta_{1}=+0.29$~mrad with $\rho=0.46$, that is, it is reported mostly as a two-qubit coupling shift and therefore stays inside the correlation sector on which the decoder prior rests, with an unabsorbed residual of $1.8\times10^{-3}$, fourteen times below the per-component shot noise. The full table is Table~\ref{tab:modelspace}. The concatenated design failed in the opposite direction---an out-of-family \emph{correlation} was absorbed as \emph{single-qubit rotations of the same size}, so the marginals survived and the correlation was deleted---and the change of frame is what removes that failure mode rather than a change of fit or of prior.

\emph{What the seven channels buy.} Because the additional references do carry identifying equations, their value is not a matter of statistical accounting at a fixed rank, and the question becomes a discrete choice among assemblies. It is settled without a trade-off. At a fixed number of shots per channel---the unit in which a QPT protocol is normally specified---the selected map lowers the worst-parameter standard deviation by a factor $63$ relative to the three-channel core, from $\sigma\sqrt{N}=24.6$ to $0.394$. At a fixed total gate budget the factor is $55$, because the four extra references cost only four gate uses per repetition; at a fixed total shot count it is $41$, because the core then receives $7/3$ times the shots per channel and is still forty times behind. There is therefore no exchange rate to fix and no rule under which the core wins, which is a cleaner statement than the three-way accounting the concatenated design required, and it is why the seven channels are the default and the core only a fallback when acquisition time is the binding constraint. Relative to the two-level concatenation used in earlier versions the change is larger still: $1.3\times10^{5}$ events and $17$ gate uses per repetition at the $0.01$~rad target, against $1.7\times10^{8}$ and $85$. Appendix~D gives the table and the wall-clock comparison.

\emph{Downstream decoder-prior test.} To verify that the recovered generator provides usable information beyond its own parameter domain, we simulate a $d=3$ rotated surface code with nine data qubits. Following the independent-block extension of Corollary 3, each of the three rows carries the same $K_\theta$, so $U_9=K_\theta^{\otimes3}$. The actual dynamics remain fully coherent. Two decoders are constructed from the Pauli-projected prior by summing all $4^9$ Pauli errors in each syndrome-and-logical class: one retains the three-qubit correlations, while the other uses exactly the same single-qubit marginals but factorizes them. Syndrome extraction is ideal and only one code-capacity round is simulated; the $14.8$-fold and $1.21$-fold gains are therefore code-capacity, single-round, ideal-syndrome statements. Two minimum extensions delimit what a deployable version must survive: a phenomenological syndrome-measurement error $p_{\rm synd}=10^{-3}$ on the eight-bit syndrome, and a cross-row $ZZ$ coupling that violates the tiled factorization. The tiled prior builds in the row factorization, so a cross-row correlation corrupts the prior in the same direction as the out-of-family absorption of Appendix~F; the gain is expected to decrease in both cases. Neither extension is simulated here. A phenomenological syndrome error needs a syndrome-level sampler and a decoder that marginalizes over syndrome faults, and a cross-row coupling breaks the $3\times$ block structure on which the $4^9$ enumeration rests; both are recorded as open extensions rather than as archived runs, and the $14.8$-fold and $1.21$-fold gains quoted above remain ideal-syndrome, single-round numbers that a deployable version has to report at the operating point it actually faces.

At the decoder point $\theta_{\rm dec}$, the unencoded average infidelity is $2.6626\times10^{-3}$, an average over the three single-qubit rotations of one block which the tiling repeats in the other two, so the per-site and the nine-qubit average coincide and the gain over the per-site figure is $14.84$. Independent-marginal decoding gives $2.2222\times10^{-4}$, whereas the correlated prior gives $1.7938\times10^{-4}$. Normalizing instead by the total nine-qubit \emph{error budget}, the probability $1-\prod_{i=1}^{9}(1-\epsilon_i)=2.37\times10^{-2}$ that at least one of the nine sites fails, which is not the nine-qubit average gate infidelity but about nine times the per-site value, gives $2.37\times10^{-2}/1.7938\times10^{-4}\approx132$; the factor three belongs to one three-qubit block, not to the nine-qubit device. The exact nine-qubit average gate infidelity of the same channel is $3.83\times10^{-2}$, giving $213$; we quote the more conservative $132$ and label it as a budget. Thus the code supplies a $14.84$-fold gain, and using the recovered correlations supplies an additional $1.239$-fold gain (a $19.3\%$ further reduction). For 16 uniformly sampled points in $\Theta$, the median QEC gain is $7.06$ (range $3.17$--$21.34$) and the median additional decoder gain is $1.065$ (maximum $1.695$).

The $1.239$ factor assumes exact knowledge of $\theta$. Because a decoder prior must in practice be built from a finite estimate, we repeated the decoding with the prior reconstructed from $\hat\theta$ sampled from the CRB covariance at the $0.01$~rad target, while the actual channel remained at $\theta$. Over $200$ draws the median gain is $1.209$, with a central $90\%$ interval of $1.173$--$1.267$, and the probability that the gain falls below $1.05$ is $1\%$. The gain halves only when the parameter error reaches about $0.057$~rad, which is more than five times the stated target and comparable to the parameter magnitudes themselves. Reaching $0.01$~rad at $\theta_{\rm dec}$ itself needs $1.55\times10^{3}$ shots, half a percent more than at $\theta_{\rm res}$: the two points are equally well conditioned, so no separate operating-point caveat is needed. The decoder result can therefore be quoted for a recovered generator rather than only for a true-generator proxy: $1.239$ assumes exact knowledge and $1.21$ is what an experiment at the stated precision would observe.

The same pipeline answers the model-error question of Appendix~F. A $Z_{1}Z_{3}$ term of amplitude $a$ is inserted symmetrically in the true channel at $\theta_{\rm dec}$, the prior is refit to the resulting fingerprint, and the gain is recomputed against the nine-qubit channel that still contains the term: $a=1$~mrad gives $1.242$ against $1.239$ in family, that is no loss at all, with the refitted generator moving by $0.83$~mrad; $2$~mrad gives $1.245$ and $5$~mrad $1.242$. The gain first falls below the in-family value at $10$~mrad, where it is $1.207$, a loss of $2.6\%$. The $1$~mrad term produces a fingerprint discrepancy $\mu=4.0\times10^{-3}$ and an unabsorbed residual $r=1.6\times10^{-3}$, the latter sixteen times below the per-component shot noise, so the term is invisible in the fit and harmless in the decoder at the same time---the combination that the concatenated design could not achieve, where the same term was silently absorbed into single-qubit offsets.

These numbers establish a proof-of-principle decoder-prior interface, not a circuit-level threshold: ancilla faults, readout errors, cross-row couplings, repeated syndromes, and the full tomographic posterior are not included. At a compensated collision the $|0\ldots0\rangle$ frame alone would assign the same fingerprint to two different physical unitaries, which propagate to different priors; in the selected fingerprint the two $|+\rangle$ references already separate the pair (Corollary~4), so the remaining requirement is a model-selection step only for accidental collisions. Full definitions and results are given in Appendix B.

\begin{table}[htbp]
\caption{Complementary noise-characterization outputs. ``Constant'' here means seven single-qubit QPT fingerprints for the fixed 11-parameter family, not a model-independent constant-cost tomography of arbitrary noise.}
\label{tab:comparison}
\centering
\scriptsize
\setlength{\tabcolsep}{3pt}
\renewcommand{\arraystretch}{1.15}
\begin{tabular}{|p{0.20\columnwidth}|p{0.31\columnwidth}|p{0.37\columnwidth}|}
\hline
Method & Primary output & Complementary role \\
\hline
Full/low-degree QPT & Process-level channel information & Broad reference and out-of-family checks; exponential cost for full QPT or structural assumptions for compression \\
\hline
ACES & Gate- and circuit-level Pauli probabilities and correlations & Direct calibration of large-circuit decoders; measures the implemented circuit rather than a unique microscopic generator \\
\hline
DGR/syndrome learning & Operational detector graph, edge/pair or hypergraph weights & Online drift tracking and mismatch correction from QEC data; does not by itself attribute a unique Hamiltonian \\
\hline
\textbf{This work} & \textbf{Coherent physical parameters in $\Theta$} & \textbf{Coupling attribution and prediction for control sequences not yet measured; transferable decoder priors} \\
\hline
\end{tabular}
\end{table}

Table~\ref{tab:comparison} emphasizes that these methods solve different inverse problems. Full and low-degree process approaches characterize channels more directly~\cite{Shabani,Crupi}; ACES estimates circuit-level Pauli noise and is already suited to large-scale decoder calibration~\cite{ACES,NoiseAware}; DGR and syndrome-only inference adapt operational decoder structures from measured QEC statistics~\cite{DGR,Takou,Cheng}. Our constant-per-fingerprint statement applies only after choosing the low-dimensional coherent family and does not imply arbitrary model-free tomography at constant cost. Conversely, a detector hypergraph can be well identified without determining a unique underlying rotation or $ZZ$ coupling, and a circuit-specific Pauli table does not predict coherent path interference after a control change.

The closest antecedent is the low-degree protocol of Ref.~\cite{Crupi}, which addresses the same noise class---coherent and correlated gate-layer noise---so the comparison is owed rather than optional. That protocol assumes a channel that is a fixed-degree polynomial in Pauli operators, i.e.\ error events touching at most $k$ qubits, and reconstructs it from random Pauli-basis preparation and measurement with an effort logarithmic in $n$ at fixed $k$, inverting the gate layer classically rather than on hardware. It is the better method asymptotically in $n$ at fixed degree, and it does not require the noise to belong to a named Hamiltonian family. The present construction makes the opposite trade: it fixes an eleven-parameter generator, pays a cost per fingerprint that is constant but needs a number of fingerprints growing linearly with $n$ (Corollary 3), and in exchange returns the generator itself rather than a channel, with the Jacobian, the conditioning, a shot budget, a complete classification of the exact ambiguities, and a decoder prior attached. Table~\ref{tab:crupi} records this boundary. Neither protocol subsumes the other, and they compose: low-degree tomography is the method of choice for large $n$ when only the channel is wanted, the polarization fingerprint when $n$ is small, parameters must be attributed to hardware sites, and the output must carry an uncertainty that survives propagation to a circuit not yet measured. A low-degree reconstruction on the same device would in fact supply the out-of-family check that a parametric fit cannot provide by itself. A depth scan with and without randomized compiling is the other competitor, and for a single known axis it is cheaper: measuring the fixed-axis fraction from one scalar needs two to three times fewer shots than the fingerprint spends on one component ($1.2\times10^{-4}$ against $2.4\times10^{-4}$~rad at $2\times10^{6}$). At depth one the two contributions are moreover exactly degenerate, so a single-depth number carries no information about their ratio at all. The comparison is therefore not fingerprint against scan but one scalar against eleven signed components at comparable cost per component; a scan cannot separate the components, attach them to sites, or say which axis a deviation lies on without assuming the axis in advance.

\begin{table}[htbp]
\caption{Division of labor with low-degree channel tomography. Entries for Ref.~\cite{Crupi} restate its published guarantees, not a re-evaluation.}
\label{tab:crupi}
\centering
\scriptsize
\setlength{\tabcolsep}{3pt}
\renewcommand{\arraystretch}{1.12}
\begin{tabular}{|p{0.17\columnwidth}|p{0.32\columnwidth}|p{0.32\columnwidth}|}
\hline
 & Low degree~\cite{Crupi} & Polarization (this work)\\
\hline
Ansatz & Fixed-degree Pauli polynomial, events on $\le k$ qubits & Named generator, $4n-1$ angles; $11$ at $n=3$\\
Cost scaling & $\log n$ at fixed degree & Constant per fingerprint; $L\gtrsim(4n-1)/12$\\
Output & Channel in the Pauli basis & Physical generator, attributable to sites\\
Uncertainty & Sample-complexity bounds & Fisher, $s_{\min}$, condition number, shot budget\\
Ambiguity & Linear inverse problem; no symmetry sector & Exact mirror sector, classified then removed\\
Downstream & Channel estimate & Pauli-projected decoder prior ($1.21\times$ at $0.01$~rad)\\
Regime & Large $n$, fixed degree & Small $n$, attribution and priors\\
\hline
\end{tabular}
\end{table}

The natural workflow is therefore hybrid. The polarization fingerprints provide an initial physical generator and its Fisher uncertainty; propagation through a proposed circuit supplies prior Pauli or amplitude-aware weights; ACES can validate the implemented circuit's Pauli model, while DGR or syndrome learning can track residual drift and out-of-model mismatch online. The surface-code test above demonstrates the first propagation step, but it neither replaces circuit-level Pauli characterization nor uses syndrome data for online refinement.

\emph{Conclusions and outlook.}---We have treated the encoding-decoding structure of quantum error correction as an information lens: coherent noise is projected onto a few logical channels and read out by single-logical-qubit process tomography, yielding a reusable physical generator rather than a static Pauli table.

The theoretical content is a reconstruction with a stated residue, not a promise. An exact rational-arithmetic certificate establishes full column rank of the $48\times11$ core Jacobian on a dense open subset of $\Theta$, hence a local inverse at every regular point, with $500$ random samples returning $s_{\min}\in[1.75,1.83]$; the same certificate covers the full map because the core rows are rows of it. What fails is injectivity in the large, and it fails at a single classified location rather than on an open set. Reflecting the chain about its middle site leaves both code channels untouched, and a $|0\ldots0\rangle$ reference degenerates where $U_3e^{-ibZ}=U_1e^{-iaZ}$, so that two regular interior points we construct share a $112$-component fingerprint while their unitaries are inequivalent. That location is classified (an eight-dimensional manifold, seven-dimensional on its mirror-fixed part) and then dissolved rather than patched: because the $|+\rangle$ reference dephases in proportion to the spectator magnetization, it separates every nontrivial mirror pair at no cost beyond preparing two spectators differently (Corollary~4). Proposition~1 upgrades the closure from an example to a class of models, the reversal being the only nontrivial permutation symmetry at any length. What remains outside it is the possibility of accidental collisions unrelated to any symmetry; the multistart search and the Jacobian sample constrain but do not exclude them, and we do not claim otherwise. The protocol also carries its model risk at the same level as its positive results, and the change of frame improves that side as well: no direction of the fifty-two out-of-family Pauli terms is amplified above its own size, and none is absorbed below $8\%$ of its response at $\theta_{\rm res}$, so the failure mode of the single-frame design---an out-of-family correlation reported as single-qubit rotations of the same size---no longer occurs. What remains is bounded rather than detected: a term that evades the fit leaves a residual that can sit below the shot noise at the operating budget, so the mandatory step is the budget of Appendices~A and~G, not a second probe.

The seven channels are chosen rather than conventional, and the choice has a measurable answer. A joint enumeration of $262\,080$ assemblies picks this set as the cheapest that retains the code diagnostic, at $1.3\times10^{5}$ events and $17$ gate uses per repetition, against $1.7\times10^{8}$ and $85$ for the two-level concatenation of earlier versions. The additional references do carry identifying equations, so their value is not a statistical accounting at fixed rank, and the three accounting rules that the concatenated design had to trade against each other now agree: the selected map is better than the three-channel core by a factor $63$ at fixed shots per channel, $41$ at fixed total shots and $55$ at fixed total gate uses, so every rule favors it and no exchange rate has to be fixed.

The numerical checks bound the inference in both directions. The equal-variance surrogate is conservative: the measurement-level Fisher information of the twelve-setting protocol is larger by a factor $1/0.72$ in the median. Drift is a scheduling constraint rather than a floor, since interleaving relaxes the tolerance by a factor $2.6$, and the blocked schedule already tolerates $2.6$~mrad per hour. Calibration is no longer the binding constraint: at the shot budget of $1.5\times10^{3}$ per component a per-gate residual of $10^{-3}$ sits an order of magnitude below the shot noise, and the worst-direction sufficient condition is $2.6\times10^{-4}$ per gate.

The downstream test shows what the generator is for. Propagating it to a distance-three surface code gives a $14.84$-fold encoding gain and a further $1.239$-fold gain from the recovered correlations, or $1.21$ when the prior is rebuilt from an estimate at the stated precision instead of from the truth. These are ideal-syndrome, single-round results; circuit-level and multi-round validation remain open.

Finally, the framework is complementary to Pauli-statistics calibration rather than competitive with it. ACES already learns circuit-level Pauli models~\cite{ACES,NoiseAware}, and DGR or syndrome-only inference adapts decoder weights from operational data~\cite{DGR,Takou}; for reducing the logical error rate of an existing circuit these are often sufficient and assume less. What a circuit-specific Pauli model does not contain is physical attribution, prediction under a changed control sequence, and an uncertainty budget tied to a generator. The natural workflow is hybrid: polarization fingerprints supply the physical generator and its Fisher uncertainty, ACES validates the implemented Pauli model, and syndrome-based methods track residual drift. Adding a non-unitary column is not a matter of enlarging the model. In a closed test in which the truth lies in the extended family, adding the matching column returns the bias to zero and recovers the parameter ($1.00\times10^{-3}$), whereas the mismatched column leaves the bias where it was ($9.5\times10^{-5}\to5.0\times10^{-5}$ for the stochastic truth, $3.4\times10^{-5}\to1.85\times10^{-5}$ for the dephasing truth, the two bias columns of Table~\ref{tab:twelfth}) and halves only the residual. The column is therefore diagnostic as well as budgetary: its fitted value is the size of the non-unitary component, and $4.2p$ is the fingerprint-norm coefficient that component sets. Pair-resolved diagnosis and held-out control sequences remain open.

\section*{Acknowledgments}
We thank Ruofan Chen and Xiansong Xu for constructive comments and discussions.

\emph{Disclosure of AI-assisted preparation.} Language editing, drafting of some analysis scripts, and cross-checks of quoted numbers against the archived outputs were assisted by a large language model. All theorems, derivations, and numerical claims were verified by the author, who takes full responsibility for the content.

\section*{Data availability}
The analysis package supporting this work contains the scripts and tabulated data behind Figs.~1--3 and Tables~I--IX: the fingerprint construction, the exact rational rank certificate, the compensated-collision construction with its 60-digit verification, the 500-point Jacobian sample, the Gauss--Newton inversion examples, the measurement-level Fisher calculation, the calibration recursion, the equal-resource Fisher comparison, the model-space scan, the non-unitary error budget, the model-error coefficient of Appendix~G, and the distance-three surface-code simulation, with their output files and the seeds listed in Appendix~C. No experimental data enter the analysis; the hardware parameters quoted in the Supplemental Material describe the device the protocol is written for and are not used in any number reported here. The package is deposited at Zenodo under the concept DOI \href{https://doi.org/\zenodoconcept}{\texttt{\zenodoconcept}}, which always resolves to the current version of the complete archive. The model-space, calibration and resource numbers reported here are reproducible from a small set of scripts in the same package: \texttt{v19\_tables.py} (every entry of the current design evaluated at $\theta_{\rm res}$ and $\theta_{\rm dec}$ in one run), \texttt{at\_res\_point.py} (the source terms of Table~\ref{tab:nuisances} and the twelfth-column test), \texttt{multistart\_v19.py} (the accidental-collision search), and \texttt{verify\_edits.py} (each number printed in this manuscript checked against the archived JSON). The domain-wide coefficients of Appendix~A, the fifteen-point model-space scan of Appendix~F, the crosstalk propagation of Appendix~B and the figure of Sec.~IV are \texttt{domain\_stability.py}, \texttt{silent\_stability.py}, \texttt{m1\_mismatch\_probe.py} and \texttt{plot\_crb\_v19.py}; the enumeration behind the design choice is in the companion package~\cite{Companion}. The Supplemental Material is deposited as a separate record under the concept DOI \href{https://doi.org/\smconcept}{\texttt{\smconcept}}, which likewise resolves to its newest version and is the DOI cited for the Supplemental Material.

\appendix
\section{Calibration-error budget for the selected fingerprint}
\label{app:calibration}
This appendix details the calibration budget quoted after Corollary~3. All entries use the operating point $\sigma_{\rm shot}=1/\sqrt{1.5\times10^{3}}=2.540\times10^{-2}$ per component at the representative parameter point; they are not domain-wide shot requirements. The quantity $\delta\eta_{\rm cal}$ is the residual per-gate calibration uncertainty, not the average gate infidelity. Representative superconducting experiments report single- and two-qubit gate errors ranging from $10^{-4}$--$10^{-3}$ on optimized devices to $10^{-3}$--several $10^{-3}$ in parallel operation~\cite{Zuchongzhi3,Marxer}. Those numbers motivate the scenarios below, but include decoherence, leakage, and stochastic noise and therefore cannot be identified directly with $\delta\eta_{\rm cal}$.

\emph{Non-unitary part.} A stochastic component of total probability $p$ at one site gives a fingerprint-space model error $\mu=4.19p$ and an unabsorbed residual $r=4.16p$; a dephasing component that reduces the coherence by $p$ gives $\mu=2.44p$ and $r=2.43p$. Both are derivatives of the fingerprint along the out-of-family direction rather than fits, $\mu=p\lVert\partial_p\mathbf M\rVert+O(p^{2})$ with $\lVert\partial_p\mathbf M\rVert=4.187$ at $\theta_{\rm res}$---the norm of the column appended to $J$, not a ratio of its in-family columns; Appendix~G gives the derivation and the block decomposition. These coefficients are domain-wide rather than point values: over the fifteen sampled points of Appendix~F the stochastic source gives $\mu/p=4.16$--$4.20$ (median $4.18$) and $r/p=4.09$--$4.19$ (median $4.13$), and the dephasing source $\mu/p=2.42$--$2.44$ (median $2.43$) and $r/p=2.40$--$2.44$ (median $2.42$), so the single-point values used here hold to $2\%$ across $\Theta$. Because $r\approx\mu$, almost the whole response is unabsorbable. At $p=10^{-3}$ the stochastic case is $\mu=4.19\times10^{-3}$ and $r=4.16\times10^{-3}$, six times below the per-component shot-noise floor $\sigma_{\rm shot}=2.54\times10^{-2}$; the component is therefore too small to bias the fit at the operating budget, and this is a statement about the size of the budget rather than about a blind spot, since $\rho=r/\mu\approx0.99$ means the response is not being absorbed by the eleven parameters at all. Raising $N$ lowers the floor and makes the term visible: $N_{\rm vis}=112/r^{2}=6.5\times10^{6}$ for $p=10^{-3}$. The two error classes are additive rather than alternative---$\delta\eta_{\rm cal}$ is what repeated calibration can remove and $\mu$ is what it cannot---and the entries below are thresholds on the first only. The latter must be obtained from repeated calibration, error-amplification sequences, and drift measurements.

\emph{Level-resolved recursion.} For a transparent scaling example take $M_0=1$ for the five product references and $M_1=6$ for the two code channels, unit sensitivities and $\kappa=1$; the code channels carry the encoding and recovery gates, the references carry none beyond the noise itself, and there is no deeper level to propagate. The reference error is $e_0=\delta$, and the code-channel error is
\[
e_1=
\begin{cases}
\sqrt6\,\delta, & p=1/2,\\
6\delta, & p=1,
\end{cases}
\]
for stochastic and fully coherent accumulation respectively. Averaged over the seven channels at the PTM-component level the combined RMS is
\[
b_p=\sqrt{\frac{5e_0^2+2e_1^2}{7}},
\]
so $C_{1/2}=1.5576$ and $C_1=3.3166$. Table~\ref{tab:calibration} evaluates the level-resolved errors and the combined RMS, and the allowance $\delta_{\max}=N^{-1/2}/C_p$. At the operating budget the allowance is $1.63\times10^{-2}$ (stochastic) or $7.66\times10^{-3}$ (coherent), so a per-gate residual of $10^{-3}$ is an order of magnitude inside it and the probe is shot-noise limited. The reverse was true of the concatenated design, where $M_2=18$ made the deepest level binding and the allowance at the same target was $1.35\times10^{-4}$; the ratio is the quantitative content of the design change. Requiring $b_p\leq0.5\,\sigma_{\rm shot}$ halves every entry in the lower panel.

\begin{table}[htbp]
\caption{Calibration budget at the representative point. Upper panel: level-resolved errors and combined fingerprint RMS for a per-gate residual $\delta$, at $\sigma_{\rm shot}=2.540\times10^{-2}$ ($N=1.5\times10^{3}$); $p=1/2$ is stochastic and $p=1$ fully coherent accumulation. Lower panel: per-gate allowance for $b_p\le\sigma_{\rm shot}$ at selected $N$, following $\delta_{\max}=N^{-1/2}/C_p$. Neither panel is a domain-wide requirement.}
\label{tab:calibration}
\centering
\scriptsize
\setlength{\tabcolsep}{1.5pt}
\renewcommand{\arraystretch}{1.0}
\begin{tabular}{cccccc}
\hline
$\delta$ & $p$ & $e_0$ & $e_1$ & $b_p$ & $b_p/\sigma$\\
\hline
$10^{-4}$ & $1/2$ & $10^{-4}$ & $2.45\times10^{-4}$ & $1.56\times10^{-4}$ & $6.1\times10^{-3}$\\
$10^{-4}$ & $1$ & $10^{-4}$ & $6.00\times10^{-4}$ & $3.32\times10^{-4}$ & $1.31\times10^{-2}$\\
$10^{-2}$ & $1/2$ & $10^{-2}$ & $2.45\times10^{-2}$ & $1.56\times10^{-2}$ & $0.613$\\
$10^{-2}$ & $1$ & $10^{-2}$ & $6.00\times10^{-2}$ & $3.32\times10^{-2}$ & $1.306$\\
\hline
$N$ & $\sigma_{\rm shot}$ & \multicolumn{2}{c}{stochastic $\delta_{\max}$} & \multicolumn{2}{c}{coherent $\delta_{\max}$}\\
\hline
$1.5\times10^{3}$ & $2.540\times10^{-2}$ & \multicolumn{2}{c}{$1.63\times10^{-2}$} & \multicolumn{2}{c}{$7.66\times10^{-3}$}\\
$10^{5}$ & $3.162\times10^{-3}$ & \multicolumn{2}{c}{$2.03\times10^{-3}$} & \multicolumn{2}{c}{$9.53\times10^{-4}$}\\
$10^{7}$ & $3.162\times10^{-4}$ & \multicolumn{2}{c}{$2.03\times10^{-4}$} & \multicolumn{2}{c}{$9.53\times10^{-5}$}\\
\hline
\end{tabular}
\end{table}

If each channel is instead required separately to satisfy $e_l\leq\sigma_{\rm shot}$ at the operating budget, the reference and code thresholds are $2.54\times10^{-2}$ and $1.04\times10^{-2}$ for stochastic accumulation, and $2.54\times10^{-2}$ and $4.23\times10^{-3}$ for coherent accumulation. The code channels are therefore the limiting contributors, but by a factor $6$ in gate count rather than the factor $18$ of the concatenated design, and neither is binding at a realistic per-gate residual.

\emph{Worst-direction bound.} A separate, deliberately conservative bound applies when coherent errors may align with the weakest parameter direction. Since each channel has $16$ PTM entries, the coherent triangle bound is
\[
\|\Delta\mathbf M\|_2\leq4(5e_0+2e_1)=68\delta .
\]
At the representative point $s_{\min}=1.7554$, so the linearized inverse satisfies
\[
\|\Delta\theta\|_2\leq\frac{\|\Delta\mathbf M\|_2}{s_{\min}}
\leq\frac{68}{1.7554}\delta .
\]
Consequently $\delta\leq2.58\times10^{-4}$ is sufficient for $\|\Delta\theta\|_2\leq0.01$~rad, and $\delta\leq1.29\times10^{-3}$ for $0.05$~rad. Written as a condition on the model error the same bound reads $\mu<s_{\min}\varepsilon=1.76\times10^{-2}$ at $\varepsilon=0.01$~rad; a stochastic component at $p=10^{-3}$ has $\mu=4.19\times10^{-3}$ and satisfies it by a factor $4.2$, and the bound is reached at $p=4.2\times10^{-3}$ for a stochastic and $7.2\times10^{-3}$ for a dephasing source. A $2\%$ readout error has $\mu=1.8\times10^{-1}$ and violates it by a factor $10$, which is why the readout threshold of the main text is the one calibration number this design actually constrains. These are worst-direction sufficient conditions, not predictions of typical experimental bias; they combine coherent gate addition, a triangle bound across channels, and inversion through $s_{\min}$. At another parameter point the bound must be rescaled by the local $s_{\min}$, which over $\Theta$ varies by $4.6\%$ and therefore barely matters here.

For nonuniform gate calibration the independent-error formula is replaced by a sum over gate types. If the code channels contain two one-qubit and four two-qubit gates each and the references one single-qubit gate each, then
\[
e_1=\sqrt{2\delta_{1Q}^2+4\delta_{2Q}^2},\qquad e_0=\delta_{1Q},
\]
and the same averaging applies. Nothing in the argument depends on the particular split; the point is that the code channels enter with the larger multiplicity and are the first place to spend calibration effort.

\emph{State preparation of the $|+\rangle$ references.} Because two of the five references prepare their spectators in $|+\rangle$, their preparation error enters the budget directly rather than through the recursion. With $\varepsilon_H$ the amplitude error of the two spectator $\pi/2$ pulses the spectator magnetization is $\langle Z\rangle\approx2\varepsilon_H$, so the unitary component left in the reference is of order $2\varepsilon_H\sin\gamma$, which for $\varepsilon_H=10^{-3}$ is $6\times10^{-5}$, four hundred times below the $2.5\times10^{-2}$ shot noise and three orders of magnitude below the $5.6\times10^{-2}$ separation that Theorem~4 establishes. Any $\varepsilon_H\ll0.1$ is adequate, so the $|+\rangle$ frame trades a gate-count cost for a preparation requirement that current devices meet by a wide margin.

\section{Decoder-prior propagation to a distance-three surface code}
\label{app:surface}

This appendix gives the numerical details behind the downstream decoder-prior test. We use the standard nine-data-qubit rotated surface code of distance three~\cite{Horsman}, with eight independent stabilizers and one logical qubit. The purpose is to test whether a physical generator recovered by the polarization fingerprints can be propagated to a different code and converted into useful decoder information. It is not a benchmark against ACES, DGR, or a fault-tolerant syndrome-extraction experiment.

\begin{figure*}[t]
\centering
\includegraphics[width=0.5\textwidth]{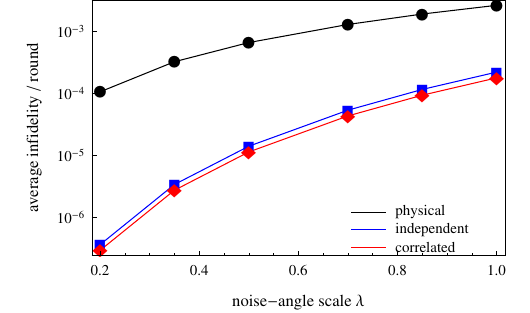}\hfill
\includegraphics[width=0.4\textwidth]{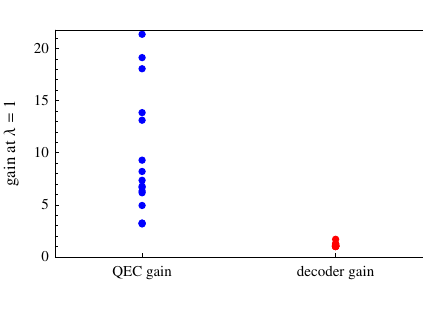}
\caption{Distance-three rotated-surface-code results under the tiled coherent-noise model. (a) Unencoded, independent-marginal-decoded, and correlated-prior-decoded average infidelities as the noise angles are scaled by $\lambda$. (b) QEC gain $r_{\rm phys}/r_{L,\rm cor}$ and additional decoder gain $r_{L,\rm ind}/r_{L,\rm cor}$ for 16 uniformly sampled parameter points at $\lambda=1$. Ideal syndrome extraction is assumed throughout.}
\label{fig:surface-gain}
\end{figure*}

\emph{Noise extension and coherent dynamics.} The three-qubit model does not uniquely determine a two-dimensional nine-qubit noise field. We therefore adopt an explicit tiled extension: the three rows of the $3\times3$ data lattice are independent blocks, and each row carries the same $K_\theta$ as in the tomography model,
\[
U_9(\theta)=K_\theta\otimes K_\theta\otimes K_\theta .
\]
There are no cross-row couplings. This is the independent-block assumption of Corollary 3, not an experimentally fitted nine-qubit noise model. The actual channel applied to the code is the coherent unitary $U_9$, without Pauli twirling. One noise layer is followed by ideal stabilizer measurement and recovery. The resulting metric is a single-round code-capacity logical average infidelity, not a circuit-level threshold.

\emph{Correlation-aware and independent-marginal decoders.} Pauli projection of each three-qubit $K_\theta$ gives 64 Pauli probabilities, and the tiled prior is their row-product. For every Pauli error $E$ on nine qubits, let $\mathrm{syn}(E)$ be its eight-bit syndrome and $\mathrm{log}(E)\in\{I,X,Z,Y\}$ its logical class modulo stabilizers. We enumerate all $4^9=262144$ errors and evaluate
\[
\begin{aligned}
P(s,\ell)&=\sum_{\substack{\mathrm{syn}(E)=s\\\mathrm{log}(E)=\ell}}p(E),\\
d(s)&=\arg\max_\ell P(s,\ell).
\end{aligned}
\]
The correlated decoder uses the full row-product $p(E)$. The independent-marginal decoder replaces it by a product of the nine single-qubit marginals of the same distribution. Thus both decoders have identical local Pauli information and include code degeneracy; their difference isolates the value of the learned correlations rather than the difference between maximum-likelihood and minimum-weight matching.

For syndrome $s$, let $A_s$ be the two-dimensional logical Kraus operator obtained after applying the selected recovery to the coherent noisy state. The logical average infidelity is evaluated exactly as
\[
r_L=\frac23\left(1-\frac14\sum_s|\mathrm{Tr}A_s|^2\right).
\]
The unencoded reference is the mean average infidelity of the three local rotations. As implementation checks, the smallest undetected logical Pauli has weight three; every single-qubit $X$, $Y$, and $Z$ error is corrected with zero residual infidelity by both decoders; and all 256 syndromes have valid recovery representatives. With a common scale $\lambda$ multiplying all noise angles, least-squares fits give exponents $1.999$ for the physical infidelity and $3.987$ and $3.986$ for the independent and correlated logical infidelities, respectively. This verifies the expected change from first-order to second-order error dominance.

\emph{Representative curve and domain sampling.} For the decoder point $\theta_{\rm dec}$
\[
\begin{aligned}
\alpha&=(0.040,-0.030,0.050),\\
\beta&=(-0.020,0.060,0.010),\\
\delta&=(0.030,-0.040,0.020),\\
(\gamma_{12},\gamma_{23})&=(0.025,-0.020),
\end{aligned}
\]
Table~\ref{tab:surface-curve} reports the noise-scale dependence and Fig.~\ref{fig:surface-gain}(a) displays it. This point is an in-domain demonstration, not an experimental fit. At $\lambda=1$, the correlated decoder reduces the average infidelity from $2.6626\times10^{-3}$ to $1.7938\times10^{-4}$. The independent-marginal decoder gives $2.2222\times10^{-4}$; hence the QEC gain is $14.84$ and the additional correlation-aware gain is $1.239$.

\begin{table}[htbp]
\caption{Representative coherent-noise curve. The final column is the additional gain from retaining correlations; the QEC gain is $r_{\rm phys}/r_{L,\rm cor}$.}
\label{tab:surface-curve}
\centering
\scriptsize
\setlength{\tabcolsep}{1.2pt}
\renewcommand{\arraystretch}{1.15}
\begin{tabular}{ccccc}
\hline
$\lambda$ & $r_{\rm ph}$ & $r_{\rm ind}$ & $r_{\rm cor}$ & gain\\
\hline
0.20 & $1.0666\times10^{-4}$ & $3.6314\times10^{-7}$ & $2.9380\times10^{-7}$ & 1.236\\
0.35 & $3.2661\times10^{-4}$ & $3.3979\times10^{-6}$ & $2.7472\times10^{-6}$ & 1.237\\
0.50 & $6.6641\times10^{-4}$ & $1.4108\times10^{-5}$ & $1.1400\times10^{-5}$ & 1.238\\
0.70 & $1.3057\times10^{-3}$ & $5.3912\times10^{-5}$ & $4.3539\times10^{-5}$ & 1.238\\
0.85 & $1.9246\times10^{-3}$ & $1.1665\times10^{-4}$ & $9.4176\times10^{-5}$ & 1.239\\
1.00 & $2.6626\times10^{-3}$ & $2.2222\times10^{-4}$ & $1.7938\times10^{-4}$ & 1.239\\
\hline
\multicolumn{5}{c}{16 random priors at $\lambda=1$: min / median / max}\\
\hline
$r_{\rm phys}/r_{L,\rm cor}$ & \multicolumn{4}{c}{$3.17$ / $7.06$ / $21.34$}\\
$r_{L,\rm ind}/r_{L,\rm cor}$ & \multicolumn{4}{c}{$1.0002$ / $1.0649$ / $1.695$}\\
\hline
\end{tabular}
\end{table}

We additionally sample 16 points uniformly from $\Theta=[-0.1,0.1]^9\times[-0.05,0.05]^2$, reconstructing both decoders at each point. The physical infidelity has median $6.90\times10^{-3}$, and the correlated-decoder logical infidelity has median $9.46\times10^{-4}$. The lower block of Table~\ref{tab:surface-curve} summarizes these gains, also shown in Fig.~\ref{fig:surface-gain}(b). Every tested point benefits from encoding. The additional value of correlations is parameter-dependent: it is small at some points and reaches a factor $1.695$ at the most favorable sampled point. No universal fixed decoder improvement is claimed.

\emph{Complementarity and limitations.} A polarization-recovered generator transfers to a code that was not used in the tomography, whereas ACES, DGR, and syndrome-only methods adapt the operational noise of the measured circuit; the two should be combined rather than substituted for one another. The gains quoted here are code-capacity, ideal-syndrome, single-round numbers, so they demonstrate the model-to-decoder interface rather than a threshold prediction.

\section{Conventions, exact certificate, and reproducibility}
\label{app:conventions}

\emph{Channel convention.} The physical unitary is
\[
K_\theta=\left(\bigotimes_{j=1}^3 e^{-i(\alpha_jX_j+\beta_jY_j+\delta_jZ_j)}\right)
e^{-i(aZ_1Z_2+bZ_2Z_3)},
\]
with $a=\gamma_{12}$ and $b=\gamma_{23}$. The parameter stacking order is
\[
\theta=(\alpha_1,\alpha_2,\alpha_3,\beta_1,\beta_2,\beta_3,
\delta_1,\delta_2,\delta_3,a,b).
\]
For each code, $W_c$ denotes the corresponding code isometry and $E_s$ the selected recovery Kraus operator. The logical Kraus operators are $A_s=W_c^\dagger E_sK_\theta W_c$, and the logical channel is $\mathcal E_{\rm L}(\rho)=\sum_s A_s\rho A_s^\dagger$. The two code channels are the ones used in the selected fingerprint; a deeper level, if one is taken, is obtained by applying the same construction to the first-level logical channel and is a deterministic function of the first-level data, which is why it cannot add an identifying equation.

For a single-qubit channel $\mathcal E$, the real PTM is
\[
(\mathcal R)_{mn}=\frac12\operatorname{Tr}\!\left[P_m\,\mathcal E(P_n)\right],
\qquad
P_m\in\{I,X,Y,Z\},
\]
with output index $m$ and input index $n$. Each $\mathcal R$ is flattened row by row, with output index varying slowest. The 112-vector is the concatenation
\[
(\Lambda_1^{\rm Z},\Lambda_2^{\rm Z},\Lambda_3^{\rm Z},
\mathcal R_X,\mathcal R_Z,\Lambda_1^{\rm X},\Lambda_3^{\rm X}),
\]
each block contributing 16 entries. This is a PTM convention, not a process-matrix convention. If a process matrix $\chi$ is reconstructed experimentally, it is converted to $\mathcal R$ before comparison. Singular values and calibration norms depend on this choice; the rank does not.

\emph{Exact local-rank certificate.} The rational certificate point is specified by the stereographic coordinates
\[
p_1=\frac{(3,2,3)}{94},\qquad
p_2=\frac{(2,-3,3)}{95},\qquad
p_3=\frac{(3,3,-2)}{96},
\]
with $t_{12}=\tan\gamma_{12}=61/1860$ and $t_{23}=\tan\gamma_{23}=53/1404$. Both couplings are Pythagorean, $1+t_{12}^2=(1861/1860)^2$ and $1+t_{23}^2=(1405/1404)^2$, and the scalar parts $q_0=(1-r^2)/(1+r^2)$ are rational in the chart below, so the exact evaluation never leaves $\mathbb Q$ and no radical appears at any intermediate step. For $p=(p_x,p_y,p_z)$ and $r^2=p_x^2+p_y^2+p_z^2$,
\[
q_0=\frac{1-r^2}{1+r^2},\qquad
(q_x,q_y,q_z)=\frac{2p}{1+r^2},
\]
and $U=q_0I-i(q_xX+q_yY+q_zZ)$. Derivatives are taken analytically in this chart, not by finite differences. The certificate was computed, as recorded below, on a $64\times11$ Jacobian of a four-block completion of the core; because rank $11$ of the completion forces rank $11$ of the $48\times11$ core, and the core rows are rows of the selected fingerprint, certifying the completion certifies the design. In that $64\times11$ Jacobian the minor with one-based rows
\[
\{1,2,3,6,8,10,11,33,34,35,37\}
\]
has a nonzero rational determinant of magnitude approximately $1.09\times10^{-49}$. Because the completion equals $DG\,DF$ for the core map $F$, with $G$ the fixed encoding-correction map that defines the four extra blocks, its rank $11$ certifies $\operatorname{rank}DF=11$ at the same point, which is the statement used by Theorem~1. All three rotation angles of the point ($0.0976$, $0.0987$, $0.0997$~rad) and both coupling angles ($0.0328$, $0.0377$~rad) lie inside $\Theta=[-0.1,0.1]^9\times[-0.05,0.05]^2$, so the certificate is obtained in the interior of the domain and requires no continuation step.

\emph{Exact compensated collision.} Set
\[
p_1=\frac{(31,24,-17)}{2000},\qquad
p_2=\frac{(-43,27,35)}{2000},
\]
\[
a=2\arctan\frac1{125},\qquad
b=-2\arctan\frac3{500},
\]
and construct
\[
\begin{aligned}
U_1&=U(q(p_1)),\qquad U_2=U(q(p_2)),\\
U_3&=U_1e^{-i(a-b)Z}.
\end{aligned}
\]
The resulting angular parameters are listed in Table~\ref{tab:collision-points}. Both $\theta$ and $S\theta$ lie strictly inside $\Theta$.

\begin{table}[htbp]
\caption{Angular parameters of the exact compensated collision in radians. The mirror $S\theta$ exchanges sites 1 and 3 and the two couplings.}
\label{tab:collision-points}
\centering
\scriptsize
\setlength{\tabcolsep}{4pt}
\begin{tabular}{ccc}
\hline
parameter & $\theta$ & $S\theta$\\
\hline
$\alpha_1$ & $0.0309953$ & $0.0316540$\\
$\beta_1$ & $0.0239963$ & $0.0231186$\\
$\delta_1$ & $-0.0169974$ & $0.0109878$\\
$\alpha_2$ & $-0.0429864$ & $-0.0429864$\\
$\beta_2$ & $0.0269914$ & $0.0269914$\\
$\delta_2$ & $0.0349889$ & $0.0349889$\\
$\alpha_3$ & $0.0316540$ & $0.0309953$\\
$\beta_3$ & $0.0231186$ & $0.0239963$\\
$\delta_3$ & $0.0109878$ & $-0.0169974$\\
$a$ & $0.0159997$ & $-0.0119999$\\
$b$ & $-0.0119999$ & $0.0159997$\\
\hline
\end{tabular}
\end{table}

The equality is algebraic. The physical reference channel is
\[
\Lambda_1^{\rm Z}(\theta)=\operatorname{Ad}_{U_1e^{-iaZ}},
\qquad
\Lambda_1^{\rm Z}(S\theta)=\operatorname{Ad}_{U_3e^{-ibZ}}.
\]
The relation $U_3=U_1e^{-i(a-b)Z}$ makes these channels identical. Every code-based channel is invariant under the reflection, so all $80$ components of $\mathbf M_Z$ agree, and the two $|+\rangle$ references differ by $5.59\times10^{-2}=\sin2(\gamma_{12}-\gamma_{23})$ at the same pair (Theorem~4), a separation that vanishes only when the pair degenerates. A high-precision evaluation with 60-digit working precision gives a maximum component difference below $10^{-55}$ within $\mathbf M_Z$ and an exactly vanishing $L_2$ difference at the reported precision.

The two points are physically inequivalent. For $K_S=K_{S\theta}$,
\[
\left|\operatorname{Tr}(K_S^\dagger K_\theta)\right|
=7.98746954<8,
\]
so the unitaries are not related by a global phase. The nontrivial eigenvalues of $K_S^\dagger K_\theta$ include $0.99373477\pm0.11176406\,i$. Independent finite-difference Jacobians of the 48-component submap at the two points both have rank $11$, with $s_{\min}=5.654\times10^{-3}$ and $6.793\times10^{-3}$, respectively. The collision therefore occurs at regular points and is not a rank-deficiency artifact.

\emph{Numerical settings and negative results.} The 500-point Jacobian study uses four batches with seeds $111$, $222$, $333$, and $444$, 125 points each, and central differences with step $10^{-5}$; parameters are drawn independently and uniformly per coordinate from $\Theta$. The exact inversion examples use seeds $777$ and $20241029$, initial perturbations of radius $0.005$, a backtracking Gauss--Newton iteration with at most 40 steps, and a stopping threshold $10^{-9}$ on the relative residual decrease. Converged residuals near $10^{-15}$ are reported as preimages, not as uniqueness proofs. Failed starts and local minima are retained in the analysis log. The surface-code calculation enumerates all $4^9$ Pauli errors and uses ideal syndrome extraction; its seeds and data are included with the analysis package.

\emph{Software and versions.} Symbolic algebra and the 60-digit collision verification were carried out in Mathematica~8.0 (the 2010 release; retained because the exact-arithmetic verification notebooks were developed and frozen on it and run unmodified on current versions) with arbitrary-precision arithmetic; the numerical linear algebra, the measurement-level Fisher calculation, the multi-start inversion, and the independent Python re-implementation of the seven channels used Python~3 with NumPy. The surface-code calculation is plain Python and needs no external solver. Interpreter and library versions, together with SHA-256 checksums of every script, are recorded in the analysis archive, so the two independent implementations can be re-run separately and compared.

\emph{Disclosure.} The disclosure of AI-assisted preparation required for this work is given in the Acknowledgments.

The analysis package accompanying this work contains executable scripts for the fingerprint, the exact rank checks, collision verification, the $500$-sample statistics, closed-loop inversion, and the calibration recursion, together with the tabulated surface-code data. A second set of scripts reproduces the supporting checks: an independent Python re-implementation of the seven channels that reproduces the collision to $2.2\times10^{-16}$ within the $|0\ldots0\rangle$ frame and the $5.59\times10^{-2}$ separation of the $|+\rangle$ frames at the same pair, together with $s_{\min}=1.751$ at both colliding points; the two-qubit marginal channel of Theorem~3; the measurement-level Fisher calculation; the sign-flip sweep over all $2047$ nontrivial flips at the certified point; the drift-bias coefficient; the multi-start collision search with domain filtering; and the $200$-draw prior-mismatch decoding study. The scripts print both successful and failed reconstructions, and the two independent implementations of the fingerprint provide a cross-check on the conventions. No proprietary data are used.

\section{Resource comparison of the core and the selected fingerprint}
\label{app:resources}

This appendix gives the resource model behind the comparison in the main text, and it also documents why that comparison no longer needs a trade-off. For each channel $c$, $J_c$ is the $16\times11$ PTM Jacobian and $A_c=J_c^TJ_c$. With $N_c$ repetitions the Gaussian-surrogate Fisher matrix is $F(\{N_c\})=\sum_cN_cA_c$, and we report the worst parameter deviation $\sigma_{\max}=[\max_j(F^{-1})_{jj}]^{1/2}$, whose absolute normalization is arbitrary because only ratios are used. Shots are $S=\sum_cN_c$ and gate uses $G=\sum_cg_cN_c$ with
\[
(g_{Z1},g_{Z2},g_{Z3},g_B,g_P,g_{X1},g_{X3})=(1,0,0,6,6,2,2),
\]
with $\sum_cg_c=17$ against $g=(1,6,6)$ and $\sum_cg_c=13$ for the core. The two $|0\ldots0\rangle$ references on sites $2$ and $3$ need no gate beyond the noise itself, the site-$1$ reference carries one preparation use, each $|+\rangle$ reference carries its two spectator $\pi/2$ pulses, and each code channel carries the six gates of encoding and recovery, so the seven entries add to $17$. These are illustrative counts of the same kind as those in the calibration recursion, not measured platform costs; a different circuit realization changes the numerical balance and not the conclusion.

\emph{The core does not win under any accounting rule.} Table~\ref{tab:resource} collects the comparison for the three rules that are normally used. At a fixed number of shots per channel---the unit in which a QPT protocol is specified---the selected map is better by a factor $24.64/0.394=63$. At a fixed total shot count the core receives $7/3$ times the repetitions per channel, which recovers a factor $\sqrt{7/3}$, and it is still worse by $41$. At a fixed total gate budget it receives $17/13$ times the repetitions, which recovers $\sqrt{17/13}$, and it is worse by $55$. The ordering is therefore the same in all three, and the trade-off that the concatenated design had to adjudicate---a statistical gain bought with $6.5$ times the gate uses---does not exist here at all.

\begin{table}[htbp]
\caption{Resource accounting for the core and the selected fingerprint at $\theta_{\rm res}$. Upper panel: worst-parameter deviation $\sigma_{\max}$ for the three accounting rules, in units of $1/\sqrt{N_c}$ for the first row and absolute for the other two; only ratios within a row are meaningful. Lower panel: repetitions per channel and total events for a $0.01$~rad worst-parameter target. The core never wins, and the margin is at least a factor $41$.}
\label{tab:resource}
\centering
\scriptsize
\setlength{\tabcolsep}{2.4pt}
\renewcommand{\arraystretch}{1.05}
\begin{tabular}{lccr}
\hline
resource rule & core (3 ch.) & selected (7 ch.) & ratio\\
\hline
equal shots per channel & $24.64$ & $0.394$ & $63$\\
equal total shots ($S=1$) & $42.7$ & $1.04$ & $41$\\
equal total gate uses ($G=1$) & $88.8$ & $1.62$ & $55$\\
\hline
repetitions per channel for $0.01$~rad & $6.07\times10^{6}$ & $1.55\times10^{3}$ & $3917$\\
total events ($12$ settings per channel) & $2.19\times10^{8}$ & $1.30\times10^{5}$ & $1680$\\
gate uses per repetition & $13$ & $17$ & $0.76$\\
settings per repetition & $36$ & $84$ & $0.43$\\
\hline
\end{tabular}
\end{table}

\emph{Why the optimization is not a free parameter.} One could try to improve the core by allocating its repetitions unevenly, and the question is whether that closes the gap. It does not, for a structural reason: the two code blocks respond to the corrected directions only at third order, so setting $N_X=N_Z=0$ makes $F$ singular. The core's bottleneck is therefore not the allocation but the absence of first-order response, and no redistribution of shots manufactures one. For the selected map the blocks are near-orthogonal and their singular values comparable, so uniform allocation is already within a few percent of the optimum. This is also why we report $\sigma_{\max}$ with uniform $N_c$ rather than an optimized frontier: the frontier is a one-parameter family in each case, and the two do not cross.

\emph{Wall-clock.} A shot count is not a time. If one repetition of channel $c$ costs $g_c\tau_g+\tau_r$, with $\tau_g$ the gate time and $\tau_r$ the readout, reset and latency per repetition, then at fixed precision
\[
\frac{W_{\rm sel}}{W_{\rm core}}=\frac{S_{\rm sel}}{S_{\rm core}}\,
\frac{\bar g_{\rm sel}x+1}{\bar g_{\rm core}x+1},
\qquad x=\frac{\tau_g}{\tau_r},
\]
with $\bar g_{\rm sel}=17/7=2.43$ and $\bar g_{\rm core}=13/3=4.33$. Because $S_{\rm sel}/S_{\rm core}=1/3920$ the selected map is faster for every $x>0$: the exchange rate that made the concatenated design win only below $x=0.043$ disappears, since the larger set is now also the cheaper one in gate uses per repetition. With readout dominating, $x\ll1$, the wall-clock ratio is $S_{\rm sel}/S_{\rm core}\cdot(1/1)$, that is three orders of magnitude; with gates dominating it is $2.43/4.33$ times that, a factor $1.8$ smaller and still decisive. The only cost that grows is the number of settings per repetition, $84$ against $36$, which enters $\tau_r$ linearly and is already included.

\emph{Calibration.} At equal per-gate precision the selected map has the smaller fingerprint-RMS coefficient, $C_{1/2}=1.558$ against $2.08$ and $C_1=3.317$ against $4.93$ for the core computed with the same recursion truncated at the code channels, because the recursion compounds the code-channel error in both cases but the selected map dilutes it over five quiet references rather than two codes. Calibration therefore does not reverse the comparison either, which is the third of the three rules under which the concatenated design had to be defended.
\section{Perturbative polarization of the two three-qubit codes}
\label{app:polarization}

Observation~1 states that the bit-flip code emphasizes the $Z$ axis and the phase-flip code the $X$ axis. Here we derive the statement to leading order, so that the mechanism does not rest on an external reference. Write a small local rotation as $\exp(-i\varepsilon\sum_iP_i)$ for a Pauli axis $P\in\{X,Z\}$, applied to all three qubits, and let $R_c$ denote the logical Pauli transfer matrix of code $c$ under the fixed standard recovery.

\emph{Commutation argument.} The bit-flip code has stabilizer generators $Z_1Z_2$ and $Z_2Z_3$. A rotation about $Z$ commutes with both generators, so it acts within each syndrome sector; on the code space all three $Z_i$ act as the logical $Z_L$, and the rotation therefore passes through as a logical $Z$ rotation of angle $3\varepsilon$. A rotation about $X$ anticommutes with a generator, is detected, and is removed by the recovery, so its effect on the logical channel appears only through the failure of the recovery at second order in the generator weight. The phase-flip code exchanges these roles.

\emph{Explicit orders.} The decay of the transverse logical component quantifies the statement. For a pure $Z$ rotation of amplitude $\varepsilon$ on each qubit, the bit-flip channel obeys
\[
1-R^{\rm bit}_{xx}=2\sin^{2}(3\varepsilon)+O(\varepsilon^{4})=18\varepsilon^{2}+O(\varepsilon^{4}),
\]
whereas the phase-flip channel, which corrects precisely this error, obeys
\[
1-R^{\rm ph}_{xx}=6\varepsilon^{4}+O(\varepsilon^{6}).
\]
Numerically, $\varepsilon=0.05$ gives $4.466\times10^{-2}$ and $3.738\times10^{-5}$, and $\varepsilon=0.1$ gives $1.747\times10^{-1}$ and $5.921\times10^{-4}$, matching the two expressions to the quoted digits. The same identities hold with $X\leftrightarrow Z$ and bit-flip $\leftrightarrow$ phase-flip interchanged. The two codes therefore respond to one and the same rotation at orders $\varepsilon^{2}$ and $\varepsilon^{4}$, a separation of $3/\varepsilon^{2}$, which is the quantitative content of the polarization statement: one code passes the rotation through at the level of the logical channel, the other removes it to higher order, and the complementary pair covers both axes.

Two caveats: the expansion is in the rotation angle and assumes the standard recovery, and it is leading order only, which is why identifiability rests on the exact rank certificate rather than on this expansion.

\section{Model space: the 52 out-of-family directions}
\label{sec:modelspace}

The eleven parameters of the model are eleven of the sixty-three nontrivial Pauli directions of a three-qubit generator ($4^{3}-1=63$): nine single-qubit rotations and the two chain couplings $Z_{1}Z_{2}$, $Z_{2}Z_{3}$. The family is therefore a subspace of the generator space, not a basis of it, and the question left open above is what the fit reports when the true noise contains one of the other fifty-two directions.

\begin{table*}[t]
\caption{The 52 Pauli directions that the eleven-parameter family does not contain, sorted by $\rho$ at $\theta_{\rm res}$. The amplification is $\max|\delta\theta|/a$ for a term $\exp(-\mathrm{i}aG)$ added on top of the model; $\rho=\|(I-JJ^{+})\delta\mathbf M\|/\|\delta\mathbf M\|$ is the fraction of the response that no choice of the eleven parameters can absorb, so $\rho\ll1$ would mean the term is absorbed silently. $\rho_{\min}$ is the smallest value over the fifteen sampled points and the last column counts how many of them fall below $0.1$. No direction is amplified above its own size anywhere, and only the first two rows are ever absorbed below $0.1$, at one and two of the fifteen points respectively. In the concatenated design of earlier versions the same scan returned $37$ amplifications above unity, a maximum of $65$, and three directions with $\rho\le0.074$ at every point.}
\label{tab:modelspace}
\scriptsize
\setlength{\tabcolsep}{2.5pt}
\begin{tabular}{@{}lrrrr@{\hspace{1.6em}}lrrrr@{}}
\toprule
direction & amp. & $\rho$ & $\rho_{\min}$ & $<0.1$ & direction & amp. & $\rho$ & $\rho_{\min}$ & $<0.1$\\
\midrule
$Y_{2}Z_{3}$           & 0.97 & 0.11 & 0.08 &  1 & $X_{2}X_{3}$           & 0.41 & 0.77 & 0.69 &  0\\
$Z_{1}Y_{2}$           & 0.97 & 0.14 & 0.05 &  2 & $Y_{1}Y_{3}$           & 0.05 & 0.78 & 0.33 &  0\\
$Z_{1}Z_{3}$           & 0.80 & 0.46 & 0.36 &  0 & $X_{1}X_{3}$           & 0.36 & 0.79 & 0.78 &  0\\
$Z_{1}Z_{2}Z_{3}$      & 0.49 & 0.52 & 0.51 &  0 & $Z_{1}Y_{2}X_{3}$      & 0.07 & 0.80 & 0.75 &  0\\
$Z_{1}X_{2}$           & 0.66 & 0.57 & 0.55 &  0 & $X_{1}Y_{2}Z_{3}$      & 0.06 & 0.84 & 0.71 &  0\\
$Z_{1}X_{2}X_{3}$      & 0.65 & 0.58 & 0.58 &  0 & $Y_{1}Z_{2}Z_{3}$      & 0.49 & 0.86 & 0.85 &  0\\
$Z_{1}X_{2}Z_{3}$      & 0.66 & 0.58 & 0.56 &  0 & $Y_{1}X_{2}X_{3}$      & 0.49 & 0.86 & 0.86 &  0\\
$X_{1}X_{2}Z_{3}$      & 0.65 & 0.58 & 0.57 &  0 & $Z_{1}Z_{2}Y_{3}$      & 0.49 & 0.86 & 0.85 &  0\\
$X_{2}Z_{3}$           & 0.64 & 0.59 & 0.53 &  0 & $X_{1}X_{2}Y_{3}$      & 0.49 & 0.86 & 0.86 &  0\\
$Y_{1}X_{3}$           & 0.52 & 0.66 & 0.65 &  0 & $Y_{2}X_{3}$           & 0.07 & 0.88 & 0.59 &  0\\
$Z_{1}X_{3}$           & 0.69 & 0.66 & 0.64 &  0 & $Y_{1}Z_{2}X_{3}$      & 0.04 & 0.89 & 0.78 &  0\\
$X_{1}Y_{3}$           & 0.51 & 0.68 & 0.65 &  0 & $Y_{1}X_{2}Z_{3}$      & 0.04 & 0.89 & 0.67 &  0\\
$X_{1}Z_{3}$           & 0.67 & 0.69 & 0.63 &  0 & $Z_{1}X_{2}Y_{3}$      & 0.04 & 0.90 & 0.70 &  0\\
$Y_{1}X_{2}$           & 0.50 & 0.69 & 0.64 &  0 & $X_{1}Z_{2}X_{3}$      & 0.03 & 0.91 & 0.85 &  0\\
$X_{2}Y_{3}$           & 0.49 & 0.70 & 0.64 &  0 & $X_{1}Y_{2}$           & 0.05 & 0.94 & 0.55 &  0\\
$Z_{1}Y_{2}Z_{3}$      & 0.97 & 0.71 & 0.69 &  0 & $Y_{2}Y_{3}$           & 0.05 & 0.95 & 0.55 &  0\\
$X_{1}X_{2}X_{3}$      & 0.50 & 0.71 & 0.70 &  0 & $Y_{1}Y_{2}$           & 0.06 & 0.95 & 0.63 &  0\\
$Y_{1}Z_{2}$           & 0.48 & 0.72 & 0.66 &  0 & $X_{1}Z_{2}Y_{3}$      & 0.02 & 0.97 & 0.78 &  0\\
$Z_{2}Y_{3}$           & 0.48 & 0.72 & 0.66 &  0 & $X_{1}Y_{2}X_{3}$      & 0.05 & 0.99 & 0.99 &  0\\
$Z_{1}Y_{3}$           & 0.47 & 0.74 & 0.65 &  0 & $Z_{1}Y_{2}Y_{3}$      & 0.05 & 1.00 & 0.99 &  0\\
$X_{1}Z_{2}Z_{3}$      & 0.42 & 0.76 & 0.74 &  0 & $Y_{1}Y_{2}Z_{3}$      & 0.06 & 1.00 & 0.99 &  0\\
$Z_{1}Z_{2}X_{3}$      & 0.41 & 0.76 & 0.75 &  0 & $X_{1}Y_{2}Y_{3}$      & 0.04 & 1.00 & 1.00 &  0\\
$X_{1}X_{2}$           & 0.42 & 0.76 & 0.70 &  0 & $Y_{1}Z_{2}Y_{3}$      & 0.04 & 1.00 & 0.99 &  0\\
$Z_{2}X_{3}$           & 0.41 & 0.76 & 0.71 &  0 & $Y_{1}Y_{2}X_{3}$      & 0.04 & 1.00 & 1.00 &  0\\
$Y_{1}Z_{3}$           & 0.45 & 0.76 & 0.65 &  0 & $Y_{1}X_{2}Y_{3}$      & 0.03 & 1.00 & 1.00 &  0\\
$X_{1}Z_{2}$           & 0.41 & 0.76 & 0.71 &  0 & $Y_{1}Y_{2}Y_{3}$      & 0.01 & 1.00 & 1.00 &  0\\
\bottomrule
\end{tabular}
\end{table*}

\emph{Construction.} Keep the model at $\theta_{\rm res}$ and add a term $\exp(-\mathrm{i}aG)$ on top of the same noise channel, one $G$ at a time, with $a=1$~mrad. The insertion is symmetric, $\exp(-\mathrm{i}aG/2)\,K\,\exp(-\mathrm{i}aG/2)$, so the added term is a generator and not an asymmetric perturbation; the channel is carried in its Kraus form, so nothing is truncated in $a$. The fingerprint is recomputed with the published pipeline (the channel and the unitary implementations agree to machine precision in the unitary limit) and the discrepancy $\delta\mathbf M=\mathbf M_{G}-\mathbf M_{0}$ is split once by the pseudo-inverse of the $112\times11$ Jacobian,
\begin{equation}
\delta\theta=J^{+}\delta\mathbf M,\qquad
\rho=\frac{\lVert(I-JJ^{+})\delta\mathbf M\rVert}{\lVert\delta\mathbf M\rVert},
\end{equation}
into the parameter shift the model reports and the fraction of the response that no choice of the eleven parameters can absorb. Both are first order in $a$, and $\delta\theta$ is the least-squares shift by construction, $J^{T}(I-JJ^{+})\delta\mathbf M=0$ to $6\times10^{-13}$.

\emph{Why the linear split is enough.} A full Gauss--Newton refit reproduces the linear shift to four digits at $1$~mrad: for a $Z_{1}Z_{3}$ term both give $0.8035$~mrad, the fit converging in four iterations to a residual of $1.8\times10^{-3}$, so at the size the table uses the split is exact rather than approximate. The second-order term appears only at larger angles, and there it reduces the amplification instead of increasing it---the same $Z_{1}Z_{3}$ term gives $0.800$ at $100$~mrad against $0.805$ linearly---so the linear amplification is an upper bound and the table is conservative in that direction.

\emph{Result.} At $\theta_{\rm res}$ the amplification $\max|\delta\theta|/a$ has median $0.42$ and maximum $0.97$ (the $Z_{1}Y_{2}$ term), and exceeds unity for none of the $52$ directions; over the fifteen sampled points the largest value anywhere is $0.99$. The residual fraction has median $0.77$ and minimum $0.11$. Nothing is silent: the smallest $\rho$ at $\theta_{\rm res}$ is $Y_{2}Z_{3}$ at $0.11$, whose response is reported as a parameter shift of the same size as the term, and the two directions that fall below $0.1$ at any point do so at one and two of the fifteen respectively, and never below $0.053$. This is the qualitative difference from the concatenated design of earlier versions, where the same scan gave $37$ amplifications above unity, a maximum of $65$, and three directions absorbed with $\rho\le0.074$ at every point. The reason is the frame and not the fit: a $|0\ldots0\rangle$ reference responds to $Z_{i}Z_{j}$ only through the products $U_ie^{-ia_iZ}$, so an out-of-family correlation could be written as single-qubit rotations of the same size; a $|+\rangle$ reference dephases at a rate set by the spectator magnetization, which is a first-order two-site response that no single-qubit rotation reproduces.

\emph{The crosstalk case.} The experimentally common $Z_{1}Z_{3}$ term illustrates the change. At $1$~mrad it is reported as $\delta_{2}=-0.80$, $\gamma_{12}=+0.51$, $\gamma_{23}=+0.47$, $\delta_{3}=+0.33$ and $\delta_{1}=+0.29$~mrad with $\rho=0.46$: most of it lands on the two couplings, so it stays inside the correlation sector on which the decoder prior rests, and the unabsorbed residual is $1.8\times10^{-3}$, fourteen times below the per-component shot noise. In the concatenated design the same term was reported as single-qubit rotations of the two outer sites with a residual $190$ times below the floor, so the correlation, which is the object the prior is built from, was deleted while the marginals survived. That is the failure mode the change of frame removes.

\emph{Three points that the table does not settle.} First, the absorbed set was never the set of next-nearest-neighbour couplings, and it is not one now: $X_{1}Z_{2}$ is a $1$--$2$ term and is loud ($\rho=0.76$), while $X_{1}X_{2}X_{3}$, the first candidate for a ``product of modeled generators'' rule, is also loud ($\rho=0.71$). Second, the two code channels and the five references are not equally exposed: restricted to the code sub-block, the two near-silent directions keep their small $\rho$, so a second code orientation would inherit them. The model-error budget of Appendix~G is therefore mandatory rather than optional, and it is the budget and not a second probe that closes the question. Third, the statement is linear and pointwise: $\rho$ is a first-order fraction, the fifteen-point scan is a sample and not a proof, and the table should be read as the measured set for this family, not as a classification of which Pauli terms are visible.

\emph{Why the near-silent directions are still bounded.} The reason is the same in both designs. If $L$ maps the block responses of one frame onto another, the response of the second frame lies in the span of the eleven generator responses whenever the eleven columns are independent, $J_B=LJ_A$ with $J_A$ of full column rank $11$, so no additional block of this family can resolve a direction that the first frame absorbs. The rank condition is what makes the argument valid, and it is certified in Appendix~C. What changes with the frame is the size of the absorbed set, from three directions at $\rho\le0.074$ to two at $\rho\ge0.053$ and only at scattered points, which is why the budget of the next section is now a bound on shot noise rather than on the fit.

\section{Non-unitary model error and its cost}
\label{sec:nonunitary}

\begin{table}[t]
\caption{Four non-unitary sources at $\theta_{\rm res}$, with $\sigma_{\rm shot}=1/\sqrt{1.5\times10^{3}}=2.54\times10^{-2}$ per component. $\delta\theta$ is the largest single-parameter shift the eleven-parameter fit reports, $r$ the unabsorbed residual, $\mu$ the norm of the model error, and $N_{\rm sat}$, $N_{\rm vis}$ the two budgets of Eq.~(\ref{eq:budgets}). A source is invisible when $r$ stays below the floor at $N_{\rm vis}$ well beyond the operating budget $1.5\times10^{3}$; every entry except the readout row satisfies that, the residual being between $6$ and $4500$ times below the floor.}
\label{tab:nuisances}
\scriptsize
\setlength{\tabcolsep}{1.5pt}
\begin{tabular}{@{}lrrrrr@{}}
\toprule
source & $\delta\theta$ & $r$ & $\mu$ & $N_{\rm sat}$ & $N_{\rm vis}$\\
\midrule
drift $1$~mrad & $1.1\times10^{-7}$ & $5.7\times10^{-6}$ & $5.7\times10^{-6}$ & $3.0\times10^{10}$ & $3.5\times10^{12}$\\
deph.\ $10^{-3}$ & $3.4\times10^{-5}$ & $2.4\times10^{-3}$ & $2.4\times10^{-3}$ & $1.7\times10^{5}$ & $1.9\times10^{7}$\\
stoch.\ $10^{-3}$ & $9.5\times10^{-5}$ & $4.2\times10^{-3}$ & $4.2\times10^{-3}$ & $5.7\times10^{4}$ & $6.5\times10^{6}$\\
readout $2\%$ & $2.9\times10^{-3}$ & $1.8\times10^{-1}$ & $1.8\times10^{-1}$ & $30$ & $3.4\times10^{3}$\\
\bottomrule
\end{tabular}
\end{table}

\begin{table}[t]
\caption{Adding a twelfth column to the eleven-parameter fit, with the truth drawn from the extended family at $\theta_{\rm res}$ and $p=10^{-3}$. ``Before'' is the largest parameter shift of the eleven-parameter fit; ``after'' is the same quantity once the column is fitted; $q$ is the fitted value of the column.}
\label{tab:twelfth}
\scriptsize
\setlength{\tabcolsep}{3pt}
\begin{tabular}{@{}llrrr@{}}
\toprule
truth & column & before & after & $q$\\
\midrule
stochastic & depolarising & $9.5\times10^{-5}$ & $0$ & $1.00\times10^{-3}$\\
stochastic & dephasing $Z_{1}$ & $9.5\times10^{-5}$ & $5.0\times10^{-5}$ & $1.33\times10^{-3}$\\
dephasing & dephasing $Z_{1}$ & $3.4\times10^{-5}$ & $0$ & $1.00\times10^{-3}$\\
dephasing & depolarising & $3.4\times10^{-5}$ & $1.85\times10^{-5}$ & $4.5\times10^{-4}$\\
\bottomrule
\end{tabular}
\end{table}

The same construction applies to errors that are not generators at all. Write $\mu=\lVert\delta\mathbf M\rVert$ for the model error in fingerprint norm, $r=\lVert(I-JJ^{+})\delta\mathbf M\rVert$ for the part the fit cannot absorb, and define the two budgets
\begin{equation}
N_{\rm sat}=\frac{1}{\mu^{2}},\qquad N_{\rm vis}=\frac{112}{r^{2}},
\label{eq:budgets}
\end{equation}
where $N_{\rm sat}$ is the number of repetitions beyond which the reported generator stops improving and $N_{\rm vis}$ the number at which the residual reaches the noise floor of the $112$-component norm, $\sqrt{112/N}$ ($1/\sqrt{N}$ per component, the two differing by $\sqrt{112}$). Between the two the reported parameter carries an error that neither more shots nor a $\chi^{2}$ test can address. Table~\ref{tab:nuisances} evaluates four sources at the operating budget of $1.5\times10^{3}$ repetitions.

\emph{Two linear laws.} For a stochastic component of total probability $p$ at one site the fingerprint moves as $\mu=4.19p$ and $r=4.16p$, with the ratios constant to $1\%$ over $p=3\times10^{-4}$, $10^{-3}$ and $3\times10^{-3}$. Throughout the paper $p$ is a per-site quantity; because the response is first order in the added term, components at several sites add, so a total stochastic budget $p_{\rm tot}$ spread evenly over $s$ sites contributes $\mu\approx4.19p_{\rm tot}/s$. The linear law is validated at $\theta_{\rm res}$; the fifteen-point scan of Appendix~F provides the domain-wide check on the family ($\mu/p=4.16$--$4.20$, median $4.18$; $r/p=4.09$--$4.19$, median $4.13$), and the archived script evaluates the coefficient at any point of $\Theta$, so the domain-wide version of this statement is a matter of running it rather than of new code. A dephasing component that reduces the coherence by $p$ gives $\mu=2.44p$ and $r=2.43p$ ($\mu/p=2.42$--$2.44$ over the same scan). Two features of these numbers matter more than their size. First, $r\approx\mu$: almost the whole response is unabsorbable, which is the quantitative form of the statement that no direction of the family is silent any longer, and it is the opposite of the concatenated design, where the stochastic coefficient was $8.4p$ but only $2.3p$ survived as residual. Second, the stochastic law is the one to remember, because $p$ is measured in every session and needs no extra assumption: it sets a floor on the model error that no protocol in this family can go below, and at $p=10^{-3}$ that floor, $4.19\times10^{-3}$, is six times below the shot-noise level $\sigma_{\rm shot}=2.54\times10^{-2}$ of the operating budget, so a device at $10^{-3}$ per-site stochastic error is limited by its shot count and not by its model error.

\emph{Where the coefficient comes from.} The fingerprint is linear in the channel, so $4.19$ is a derivative rather than a fit. Writing $\mathcal D_p$ for the depolarizing channel of total probability $p$ on site~1,
\begin{equation}
\begin{split}
\mu(p)&=\bigl\lVert\mathbf M(K_\theta\circ\mathcal D_p)-\mathbf M(K_\theta)\bigr\rVert\\
&=p\,\lVert\partial_p\mathbf M\rVert+O(p^2),
\end{split}
\label{eq:mu-derivative}
\end{equation}
where $\lVert\partial_p\mathbf M\rVert=4.187$ at $\theta_{\rm res}$ and the quadratic term is at the $0.1\%$ level for $p=10^{-3}$. Because $p$ is not a parameter of the family, this is not a ratio of two Jacobian columns: it is the norm of the fingerprint derivative along an out-of-family direction, to be compared with column norms of $2.85$--$4.88$ per rad for the eleven angles. The response is carried by exactly the four channels that touch the depolarized site: $\lVert\partial_p\Lambda_1^{\rm Z}\rVert=2.31$ and $\lVert\partial_p\Lambda_1^{\rm X}\rVert=2.31$ for the two frames on site~1, and $1.85$ for each code channel, while the references on sites $2$ and $3$ are exactly blind to it, as they must be; the sum of squares recovers $4.187$. This block decomposition is the quantitative form of the statement that the budget is set by the number of independent first-order views of a site, not by the depth of the circuit. For scale, $p=10^{-3}$ displaces the fingerprint by $4.2\times10^{-3}$, which is the size of the response to a $1$~mrad single-qubit rotation ($2.9$--$4.9\times10^{-3}$).

The same numbers explain why the residual is a poor alarm. At $p=10^{-3}$ both the model error and its unabsorbed part are a sixth of the shot-noise floor ($\mu/\sigma_{\rm floor}=0.165$ and $r/\sigma_{\rm floor}=0.164$), so neither the fit nor a $\chi^{2}$ test would flag the term at the operating budget, and the term becomes visible only at $N\ge N_{\rm vis}=6.5\times10^{6}$. The reason is the size of the shot budget and not absorption: $\rho=r/\mu\approx0.99$, so the response is not being written into the eleven parameters at all. A device whose per-site stochastic error is $10^{-2}$ rather than $10^{-3}$ would instead have $r/\sigma_{\rm floor}=1.6$, and the alarm would fire.

\emph{The worst-direction bound, evaluated.} The sufficient condition derived above, $\mu<s_{\min}\varepsilon$, allows $\mu<1.76\times10^{-2}$ at $\varepsilon=0.01$~rad. Measured: a stochastic component at $p=10^{-3}$ satisfies it by a factor $4.2$ and a dephasing component by $7.2$, zero-mean drift at $5.7\times10^{-6}$ by three orders of magnitude, and only a $2\%$ readout error exceeds it, by a factor $10.4$. The bound is therefore a statement about the worst direction and, for this design, one that realistic non-unitary error meets; the readout row is the exception, and it is why readout calibration is the one number the budget actually constrains. What a session can certify is still the pair $|\delta\theta|<\varepsilon$ and $N<N_{\rm vis}$, reported per source.

\emph{The bias does not follow the signal.} Halving all eleven parameters and repeating the three perturbations, the largest reported shift grows for two of them: by $2.2$ for readout ($0.155\to0.342$~rad), by $4.0$ for dephasing ($1.5\times10^{-2}\to6.1\times10^{-2}$~rad), and stays put for the $Z_{1}Z_{3}$ crosstalk ($1.08\to1.04\times10^{-3}$~rad). A ratio of one would mean an additive floor, a ratio of two a purely multiplicative error; there is no case in which a smaller signal is safer, and two in which it is worse. The origin is the one point where the comparison fails, and not for a reason that reflects on the map: halving zero leaves zero, so with no noise to perturb the fingerprint is the identity channel for every block and there is no signal to halve.

\emph{The twelfth column.} The shifts above look like a precision ceiling. They are not: they are the price of a missing column. Adding one non-unitary nuisance parameter and refitting in closed tests in which the truth lies in the extended family,

a matching column returns the shift to zero and recovers the parameter, while a mismatched one only halves it, $3.4\times10^{-5}\to1.85\times10^{-5}$ for the dephasing truth, and reduces the residual to $1.5\times10^{-3}$ rather than to zero. Two consequences. The column is a diagnostic: its fitted value is the size of the non-unitary component, and its uncertainty is what $N_{\rm sat}$ and $N_{\rm vis}$ are computed from. And the residual is not a substitute for it: a mismatched column moves the shift down by $45\%$ and the residual by $37\%$, in the same direction as the right column but without closing either, so a $\chi^{2}$ improvement is not evidence that the right column was added.

\emph{Why a single depth cannot see the fixed-axis fraction.} For a depth-$d$ experiment with per-gate coherent amplitude $A$ and stochastic rate $B^{2}$, the average infidelity is $r(d)=(2/3)(d^{2}A^{2}+dB^{2})$ without randomized compiling and $r(d)=(2/3)d(A^{2}+B^{2})$ with it. At $d=1$ the two expressions are identical, not merely close: $1.333\times10^{-4}$ in both cases, so a single-depth number carries no information about the ratio $w=A^{2}/(A^{2}+B^{2})$. At $d=3$ the ratio of the two is $2.00$ and at $d=10$ it is $5.50$, so the information appears only with depth. Measuring $\gamma_{12}$ three ways at $A=10$~mrad and $2\times10^{6}$ repetitions gives $\sigma_{A}=1.2\times10^{-4}$~rad from the depth slope, $7.8\times10^{-5}$~rad from a randomized-compiling pair at $d=10$, and $2.4\times10^{-4}$~rad from the fingerprint component: two to three times fewer shots for the scan. The fingerprint's return is not fewer shots but eleven signed components at comparable cost per component, whose standard deviations at $2\times10^{6}$ span a factor twelve ($7.8\times10^{-5}$ to $9.4\times10^{-4}$~rad), and the ability to say which axis a deviation lies on without assuming it.

\end{document}